\documentclass[%
 reprint,
 amsmath,amssymb,
 aps,
]{revtex4-2}

\usepackage{graphicx}
\usepackage{dcolumn}
\usepackage{bm}

\usepackage[utf8]{inputenc} 
\usepackage[T1]{fontenc}    
\usepackage{hyperref}       
\usepackage{url}            
\usepackage{booktabs}       
\usepackage{nicefrac}       
\usepackage{microtype}      
\usepackage{natbib}
\usepackage{nccmath}
\usepackage[english]{babel} 
\usepackage{amsmath,amsfonts,amsthm} 
\usepackage{float}
\usepackage{blindtext}
\usepackage[sc]{mathpazo}
\usepackage{lettrine} 
\usepackage{paralist} 
\usepackage{titlesec}

\usepackage{tabularx}
\usepackage{mathtools}
\usepackage{mathptmx}
\usepackage[caption=false]{subfig}
\usepackage[font=small,justification=justified,skip=2pt]{caption}

\titlespacing*{\subsection}
  {0pt}{.5\baselineskip}{.5\baselineskip}
\titlespacing*{\subsubsection}
  {0pt}{.5\baselineskip}{.5\baselineskip}
\titlespacing*{\section}
  {0pt}{.5\baselineskip}{.5\baselineskip}

\titleformat*{\section}{\normalsize\centering\bf}
\titleformat*{\subsection}{\normalsize\centering\bf}
\titleformat*{\subsubsection}{\normalsize\centering\bf}

\usepackage{xcolor}

\makeatletter
\renewcommand*{\fnum@figure}{{\normalfont\bfseries \figurename~\thefigure}}
\makeatother

\begin{document}

\title{The role of gap junctions and clustered connectivity in emergent synchronisation patterns of inhibitory neuronal networks}

\author{Hélène Todd$^1$}
\author{Mathieu Desroches$^{2,3}$}
\author{Alex Cayco-Gajic$^{1,*}$}
\author{Boris Gutkin$^{1,*}$}
\affiliation{
$^{1}$  \small Group for Neural Theory, Laboratoire des Neurosciences Cognitives et Computationnelles, \textit Ecole Normale Supérieure PSL University, Paris, France. \\
$^{2}$ \small MathNeuro Team, \textit Inria Branch of the University of Montpellier, Montpellier, France. \\
$^{3}$ MCEN group, Basque Center for Applied Mathematics (BCAM), Bilbao, Spain. \\
\small * Equal Contribution
} 

\date{\today}


\begin{abstract}
Inhibitory interneurons, ubiquitous in the central nervous system, form networks connected through both chemical synapses and gap junctions. These networks are essential for regulating the activity of principal neurons, especially by inducing temporally patterned dynamic states. Here, we aim to understand the dynamic mechanisms that allow for synchronisation to arise in networks of electrically and chemically coupled interneurons. To this end, we use the exact mean-field reduction to derive a neural mass model for both homogeneous and clustered networks. We first analyse a single population of neurons to understand how the two couplings interact with one another. We demonstrate that the network transitions from an asynchronous to a synchronous regime either by increasing the strength of the gap junction connectivity or the strength of the background input current. Conversely, the strength of inhibitory synapses affects the population firing rate, suggesting that electrical and chemical coupling strengths act as complementary mechanisms by which networks can tune synchronous oscillatory behaviour. Next, inspired by the existence of multiple interconnected interneuron subtypes in the cerebellum, we analyse networks consisting of two clusters of cell types defined by differing chemical versus electrical coupling strengths. We show that breaking the electrical and chemical coupling symmetry between these clusters induces bistability, so that a transient external input can switch the network between synchronous and asynchronous firing. Together, our results show the variety of cell-intrinsic and network properties that contribute to synchronisation of interneuronal networks with multiple types of coupling.
\end{abstract}

\maketitle

\section{INTRODUCTION}

Inhibitory interneurons are abundant in the mammalian brain and can be found in various regions such as the neocortex, thalamus, and cerebellum, where they are thought to play a crucial role in transmitting and regulating activity in neuronal networks. They have also sparked much interest due to their connectivity properties, as interneurons have been shown to form interconnected networks through both inhibitory chemical synapses and electrical gap junctions \cite{galarreta2001, kim2021}. Experimental work has found that this combination of gap junctions with chemical synapses tends to enhance synchronous spiking in pairs of neurons \cite{holzbecher2018} and in networks \cite{bartos2007, traub2001}. This is thought to occur due to the presence of gap junction induced  spikelets, or transient depolarisations in the postysnaptic neuron when the presynaptic neuron fires \cite{kim2021}. This mechanism is generally thought to enhance spiking activity over a narrow temporal window for spiking \cite{hoehne2020}, thus potentially playing an important role in the emergent synchronisation and recruitment of principal neurons.  

Yet, challenging the intuitive view that gap junctions promote synchrony by depolarizing their neighbours, several experimental studies have demonstrated examples in which they instead \emph{desynchronise} spiking activity \citep{vervaeke2010, connors2017, hurkey2023}. These findings are supported by several studies in the modelling literature showing that electrical coupling can lead to synchronous, asynchronous, or antisynchronous firing in networks of integrate-and-fire (IF) neurons. Indeed, previous work has demonstrated that in minimal circuits \cite{rinzel2003, chow2000}, the presence of gap junctions allows for the emergence of a bistable regime, in which neurons fire either in antiphase or in synchrony, depending on their initial phase and the magnitude of the perturbation introduced within the system. These results have been extended to larger networks through classical mean-field models, showing qualitatively similar results \cite{ostojic2009, lau2010, pfeuty2007}. Still, the precise dynamical mechanisms through which the interplay of non-instantaneous chemical synapses and gap junctions drive synchronous or asynchronous spiking remain to be completely integrated within a theoretical framework.

Also important is the question of how interneuronal synchrony may support the functional role of the circuits in which they are embedded. For example, molecular layer interneurons (MLIs) in the cerebellum form coupled networks that are believed to induce and shape synchronous activity in their target Purkinje neurons that are a key site of motor learning and output of the cerebellar cortex \cite{brown2019, blot2016, mittmann2005}. Therefore, changes in molecular layer interneuron synchrony may be of functional significance to motor function by gating cerebellar output \cite{person2012}. 

In contrast to the generic homogeneous networks usually considered in modelling studies, MLIs form clustered networks, with varying levels of electrical and chemical coupling depending on cell type \cite{kozareva2021, kim2021}. However, the mechanisms by which such non-homogeneous networks may entrain synchrony in neighbouring clusters is not yet well understood. 

A promising framework to characterize the dynamical mechanisms determining synchrony is the neural mass model.  Over the last decade, a new generation of neural mass models have emerged, making it possible to perform \emph{exact} mean-field reductions of networks of heterogeneous all-to-all coupled quadratic integrate-and-fire (QIF) neurons \cite{montbrio2015}. This approach presents many advantages; first, the exact reduction enables the use of mathematical tools from dynamical systems to analyse the network. Second, these new neural mass models can incorporate various cell properties, such as electrical synapses \cite{montbrio2019}, slow chemical synapses \cite{montbrio2017, coombes2021}, and asymmetric reset values \cite{montbrio2020}, as well as sparse connectivity \cite{divolo2020, avitabile2022}, providing a rich framework to investigate synchronisation in a more biologically realistic setting.

In this contribution we aim to render a unified analysis for the impact of electrical gap junction coupling in both homogeneous and clustered inhibitory spiking networks. Specifically, we focus on characterising the neural and circuit mechanisms that allow for synchronous oscillations to emerge. Previous work has explored individually the effects of chemical coupling versus electrical coupling \cite{montbrio2019}, slow synapses \cite{montbrio2017}, asymmetric spike-resets \cite{montbrio2020} and fully symmetric clustered networks \cite{ferrara2022} within the neural mass model framework. However, to our knowledge, no study has extended these results to investigate the dynamics within a clustered network structure, nor to the concurrent addition of gap junction coupling and slow inhibitory synapses. In section \ref{sec:2}, we present the single population model, taking into account the biological aspects mentioned in the previous paragraph. We then establish the theoretical background for understanding the dynamics in a single population of all-to-all coupled neurons. We find that electrical coupling plays a crucial role for oscillations to emerge and that the onset of these oscillations is advanced by external input, or delayed by the addition of high heterogeneity or slow synapses. Finally, in section \ref{sec:3}, we present the major result of this paper: we examine two non-identical connected clusters (where one is dominated by inhibitory chemical synapses and the other by gap junctions): we find that in such a configuration, a coupling-dependent bistable regime emerges where synchrony and asynchrony co-exist. We conclude by showing that the injection of a brief external pulse current in the appropriate cluster provides a viable mechanism for switching between asynchronous and synchronous states. \\ 

\section{SINGLE POPULATION} \label{sec:2}

\subsection{The model} 

\subsubsection{Microscopic level}

Let us consider an all-to-all coupled network of $N$ quadratic integrate-and-fire (QIF) neurons with electrical and chemical synapses \cite{montbrio2019}. Each neuron's dynamics are driven by the following ordinary differential equation (ODE): 
\begin{equation} \label{eq:qif}
\tau_m \dot{v}_j(t) = v_j(t)^2 + J \tau_m s(t) + g(\bar{v}(t) - v_j(t)) + \eta_j + I(t), \end{equation}
where $v_j(t)$ is the voltage of neuron $j$, $\bar{v}(t)=\frac{1}{N}\sum_{i=1}^N v_i(t)$ the mean voltage across the network and $\tau_m$ is the membrane time constant. The action potential is modelled through a discrete reset rule such that when the membrane potential reaches the threshold $v_{th}>0$, it resets to $v_r<0$ and the neuron is considered to have spiked. As introduced in \cite{montbrio2020}, we also allow for the reset to be non-symmetric, i.e., $v_r$ is not necessarily equal to $-v_{th}$. The external input to neuron $j$ is split into two terms: the time-varying homogeneous input current $I(t)$ which is the same for all cells, and the static heterogeneous input  $\eta_j$ which is assumed to be distributed according to the Lorentzian (or Cauchy) distribution:
$$ P(\eta_j) = \frac{1}{\pi}\frac{\Delta}{(\eta_j - \bar{\eta})^2 + \Delta^2} \cdot $$ 

\noindent The instantaneous network firing rate is given by:
$$ r(t) = \lim_{\tau_s \to 0} \frac{1}{N} \frac{1}{\tau_s} \sum_{j=1}^N \sum_k \int_{t-\tau_s}^{t} \delta(t'-t_j^k)dt', $$
where $t_j^k$ is the time of the $k$th spike of neuron $j$ and $\delta(t)$ the Dirac delta function. \\

\noindent In this homogeneous all-to-all network, chemical synapses are modelled by:
$$J \tau_m s(t),$$ 
where $J$ is the strength of the chemical coupling and the synaptic activation term $s(t)$, introduced in \cite{montbrio2017}, is a filtered version of the instantaneous network firing rate across the population $r(t)$, i.e.
$$ \tau_d \dot{s}(t) = -s(t) + r(t), $$
with time constant $\tau_d$ controlling the speed of the synaptic dynamics. In what follows, we will consider only inhibitory synapses, and therefore set $J \leq 0.$ Electrical synapses, or gap junctions, are modelled by the Ohmic term:
$$g(\bar{v} - v_j(t)),$$ 
where $g$ is the strength of the electrical coupling. Later, we will consider simulations in which two clusters of neurons are connected by different chemical and electrical coupling strengths. This will allow us to extend our one-population analyses to clustered networks.

\subsubsection{Macroscopic level} 

Following \cite{montbrio2015, montbrio2017, montbrio2019, montbrio2020}, we analytically derive the dynamics of the mean-field firing rate and voltage from the microscopic equations using the Ott-Antonsen ansatz (see Appendix V A and \cite{ott2008, montbrio2015} for more details):
\begin{equation} \label{eq:mean_field}
\begin{split}
\tau_m\dot{r}(t) =& \ \dfrac{\Delta}{\tau_m\pi} + 2r(t)v(t) - 2\tau_m\ln(a)r(t)^2 - gr(t), \\[0.1em]
\tau_m\dot{v}(t) =& \ v(t)^2 + \bar{\eta} + J\tau_m s(t) + I + \dfrac{\Delta\ln(a)}{\pi} \\
&-(\ln(a)^2 + \pi^2)(\tau_mr(t))^2, \\[0.2em]
\tau_d\dot{s}(t) =& -s(t) + r(t), 
\end{split}
\end{equation}
where $a \coloneqq |v_{th}/v_r|$ is an asymmetry parameter corresponding to the proportion between the voltage threshold $v_{th}$ and the reset value $v_r$ in absolute value.

\subsection{Existence of two dynamical regimes}

\begin{figure}[H]
\centering
\includegraphics[width=\linewidth]{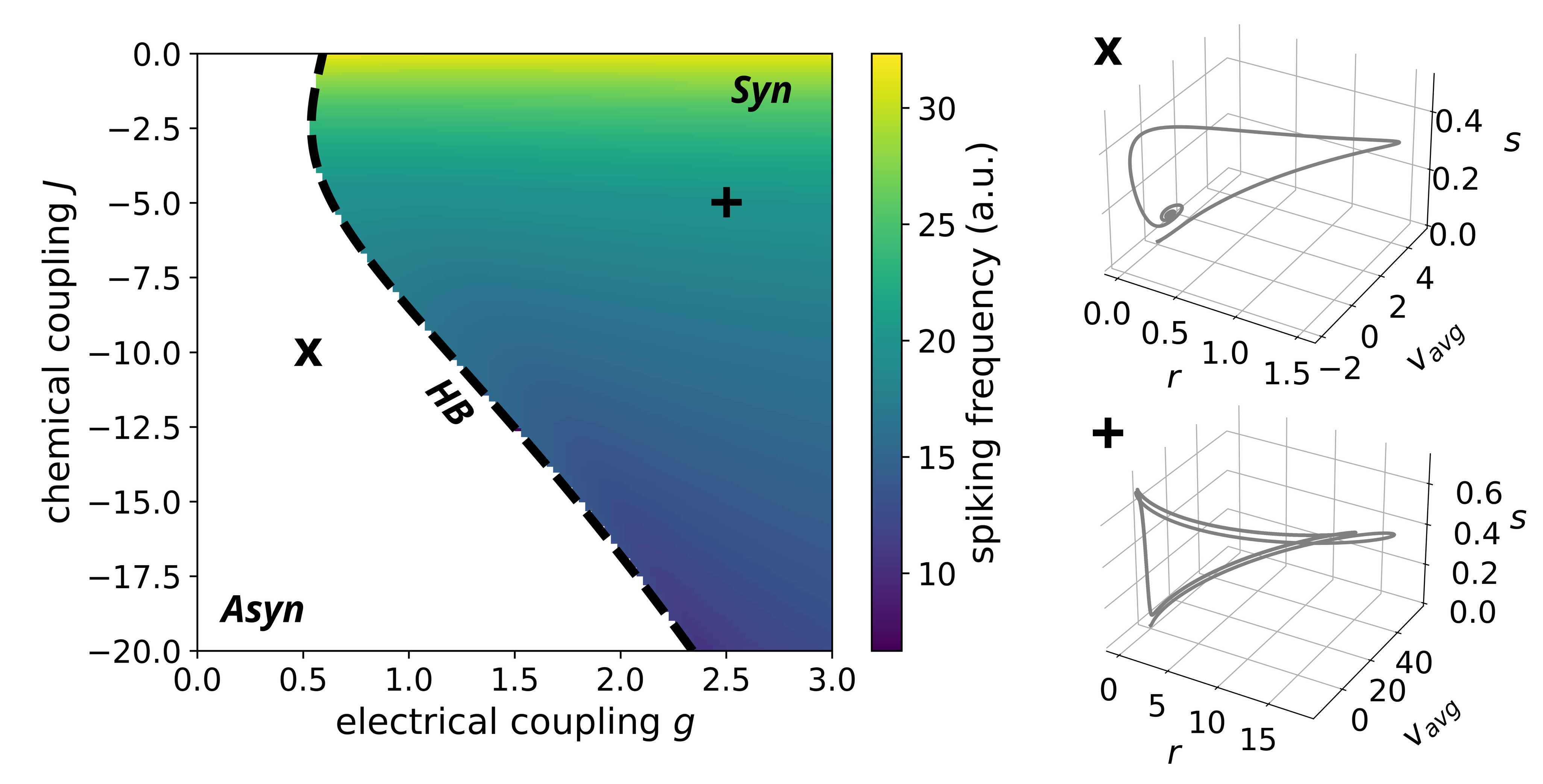}
\caption{Emergence of two dynamical regimes within the network. Left: the bifurcation diagram illustrates the roles of chemical and electrical coupling $J$ and $g$ on the network firing rate. Two distinct regimes emerge: an asynchronous (Asyn) regime, in which stable foci (SF) are predominant, and a synchronous (Syn) regime, consisting of stable limit cycles (LCs). In the (Syn) regime, the colourmap represents the frequency of the network firing rate oscillations. Right: three-dimensional plots of the dynamics of the system: example of a SF in the Asyn regime and a LC in the Syn regime. Parameters: $\bar{\eta}=1, \Delta=0.3, I=0, \tau_m=\tau_d=1, a=1.$ }\label{fig:bifs}
\end{figure}

In order to study the effects of electrical versus chemical coupling on the network, we start by considering the dynamics of the reduced model given by Eq. \ref{eq:mean_field} for fixed parameters $\bar{\eta}, \Delta, I, \tau_m, \tau_d$ and $a$ as we vary the strength of both the electrical and the inhibitory chemical coupling. We find that the system goes through a Hopf bifurcation (HB) and thus exhibits two separate regimes depending on the coupling strengths. These regimes consist of an asynchronous (Asyn) regime and a synchronous (Syn) regime (Fig. \ref{fig:bifs}). 

In the asynchronous regime, numerical simulations show that the network exhibits damped oscillations that eventually converge towards a fixed point attractor, i.e. the equilibrium is a stable focus; this behaviour reflects neural asynchrony. In the synchronous regime, the system exhibits sustained oscillations across the network, i.e. the equilibrium is a limit cycle; this behaviour reflects neural synchronisation at the population level. We find that electrical coupling is necessary for oscillations to emerge, and chemical coupling changes the global firing rate frequency; increasing the inhibitory chemical coupling decreases the spiking frequency (Fig. \ref{fig:bifs}).

In what follows, we will examine how the parameters of the system affect the dynamics. We begin by studying the effect of the asymmetry $a$ between the reset and the threshold values on the system (section \ref{sub:asymm}), followed by the synaptic time constant $\tau_d$ (section \ref{sub:time_const}), the input and mean excitability parameters $I$ and $\bar{\eta}$ (section \ref{sub:input}) and finally the heterogeneity parameter $\Delta$ (section \ref{sub:hetero}). \\

\subsection{Increased threshold advances oscillations} \label{sub:asymm}
\begin{figure}[H]
    \centering
    \includegraphics[width=\linewidth]{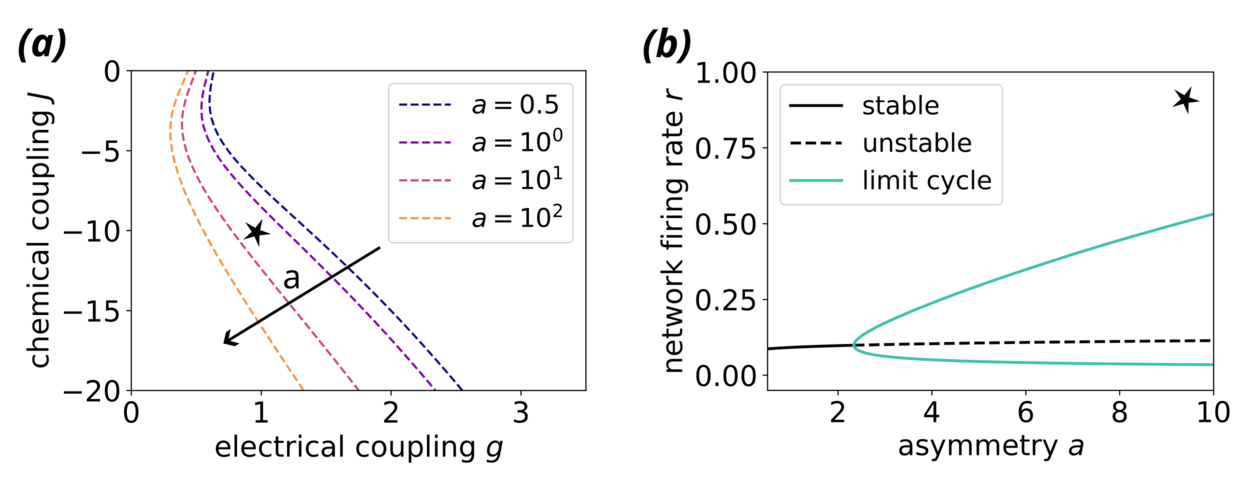}
    \caption{High threshold values $v_{th}$ advance oscillations. \textbf{(a)} Bifurcation diagram of chemical $J$ versus electrical $g$ coupling strengths for different values of asymmetry $a.$ \textbf{(b)} Bifurcation diagram showing how the asymmetry $a$ affects the network firing rate $r.$ For fixed coupling strengths $J=-5$ and $g=1$, oscillatory behaviour starts at $a \approx 2.3.$ To the right of $a \approx 2.3$, the minimum and maximum values of the oscillations for the network firing rate are plotted in green. Parameters: $\bar{\eta}=1, \Delta=0.3, I=0, \tau_m=\tau_d=1.$}
    \label{fig:bif_a}
\end{figure}

We begin by fixing $v_r=-100$ and increase $v_{th}$ in order to study how the asymmetry parameter $a$ affects the dynamics. We observe that increasing $a$ leads to a shift in the Hopf bifurcation point to lower values of $g$ (Fig. \ref{fig:bif_a}a), thus qualitatively preserving the existence of the (Asyn) and (Syn) regimes. We further illustrate this point by fixing coupling strengths and increasing $a$; the equilibrium of the system, initially stable, becomes unstable for a critical value $a$ at which it gives rise to a stable limit cycle (Fig. \ref{fig:bif_a}b).

Following the framework laid out in  \cite{montbrio2020}, we can derive the mean-field equations in the case of slow synapses. Indeed, by rewriting the voltage $v(t)$ as the voltage with symmetric resetting $v_s(t)$ plus a correction term $\tau_m \ln(a) r(t),$ the system of equations (\ref{eq:mean_field}) can be written as follows:

\begin{equation} \label{eq:mean_field_sym}
    \begin{split}
    \tau_m \dot{r} (t) =&  \ \frac{\Delta}{\tau_m \pi} + 2r(t)v_s(t)-gr(t), \\[.5em]
    \tau_m \dot{v}_s(t) =& \ v_s(t)^2 + \bar{\eta} + J\tau_m s(t) + I - \tau_m^2 \pi^2 r(t)^2 \\
    & + g\tau_m \ln(a) r(t), \\[.5em]
    \tau_d \dot{s}(t) =& \ -s(t) + r(t).
    \end{split}
\end{equation}

From the system of equations (\ref{eq:mean_field_sym}), it becomes clear that just as in the two-dimensional case, the effects of $a$ on the dynamics is closely linked to the electrical coupling term $g$.  Since parameter $a$ has no effect on the preservation of the qualitative network behaviour, we will set $a=1$. \\

\subsection{Slow synapses hinder synchronisation} \label{sub:time_const}  
 In biological neurons, chemical synapses are slower than gap junctions; we therefore explored how the chemical synaptic time constant impacts the synchronous oscillatory behaviour of the network, using  numerical continuation via AUTO. We start by noting that the Hopf bifurcation is dependent not on $\tau_d$ alone but rather on the ratio $\tau_m/\tau_d$; we therefore set $\tau_m=1$ without loss of generality in what follows, allowing us to focus on the synaptic time constant $\tau_d.$

\begin{figure}[H]
\centering
\includegraphics[width=\linewidth]{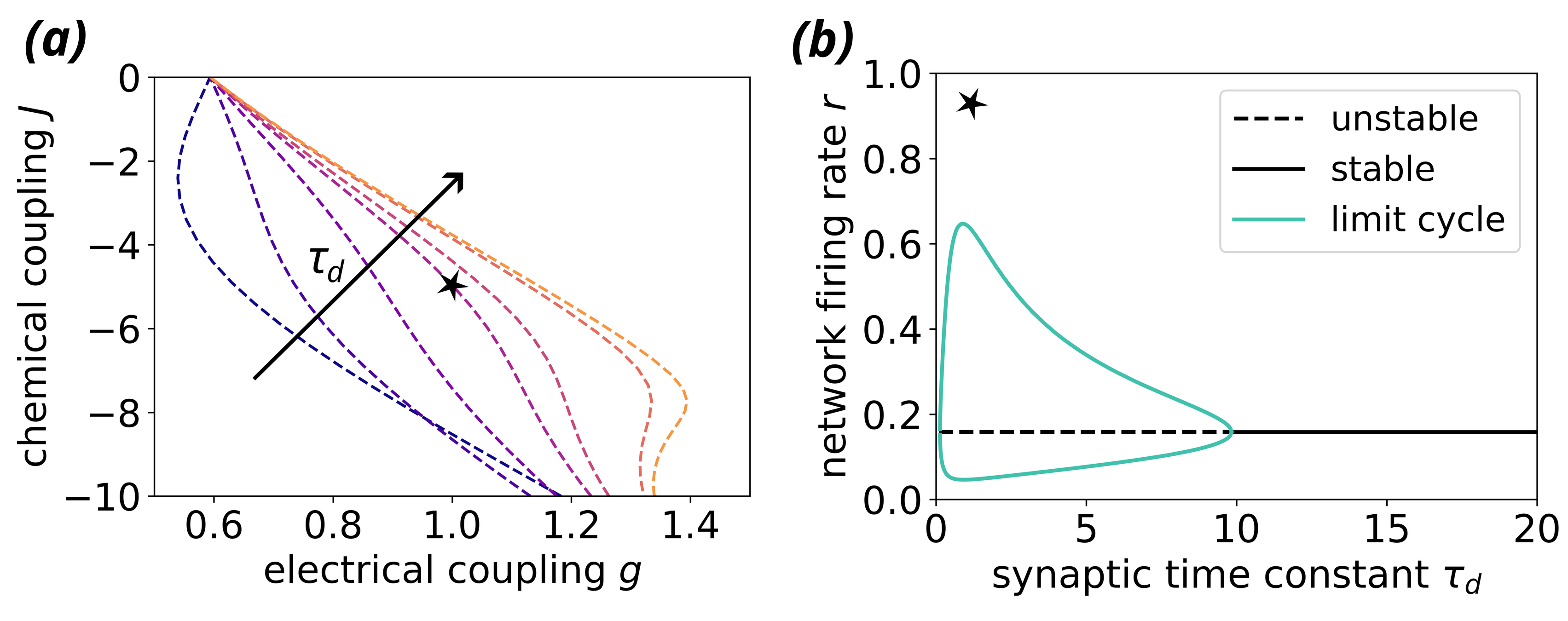}
\caption{Slow synapses hinder synchronisation. \textbf{(a)} Bifurcation diagram of chemical $J$ versus electrical $g$ coupling strengths for different values of synaptic time constants $\tau_d.$ Dashed lines represent Hopf bifurcation curves. From left to right: $\tau_d = 1, 2.5, 5, 10, 15, 50$ and $100.$ \textbf{(b)} Bifurcation diagram showing how the synaptic time constant $\tau_d$ affects the network firing rate $r$. For fixed coupling strengths $J=-5$ and $g=1$, oscillatory behaviour stops at $\tau_d=10.$ Parameters: $\bar{\eta}=1,\Delta=0.3, \tau_m=1$.}
\label{fig:bif_syn}
\end{figure}

 We find that the inhibitory synaptic time constant $\tau_d$ plays a role in the network's dynamics by changing the stability of the equilibria, thus shifting the Hopf bifurcation to higher coupling strength values as $\tau_d$ increases. Indeed, slow synapses decrease the network firing rate (Fig. \ref{fig:bif_syn}b). Sufficiently slow synapses can even prevent oscillatory behaviour for a fixed pair of coupling strengths (Fig. \ref{fig:bif_syn}a,b). Therefore, in order for there to be oscillations within the system for a wider range of coupling strengths $g$ and $J$, the synaptic time constant $\tau_d$ must not be too large. In what follows, we set $\tau_d=1$. 
 
\subsection{Increased inputs advance oscillations}  \label{sub:input}

Interneurons receive inputs from excitatory neurons; we therefore examine how introducing a positive, external input current into the network influences its behaviour. Let us begin by noting that in Eq. \ref{eq:mean_field}, the mean heterogeneity $\bar{\eta}$ and homogeneous input current $I$ play the same role. We can therefore refer to the parameters $\bar{\eta}$ and $I$ interchangeably when considering the behaviour at the level of the whole network. For the remainder of the paper, we will consider changing $I$ as $\bar{\eta}$ is assumed to be a time-independent property of the neurons themselves. 

\begin{figure}[H]
\centering
\includegraphics[width=\linewidth]{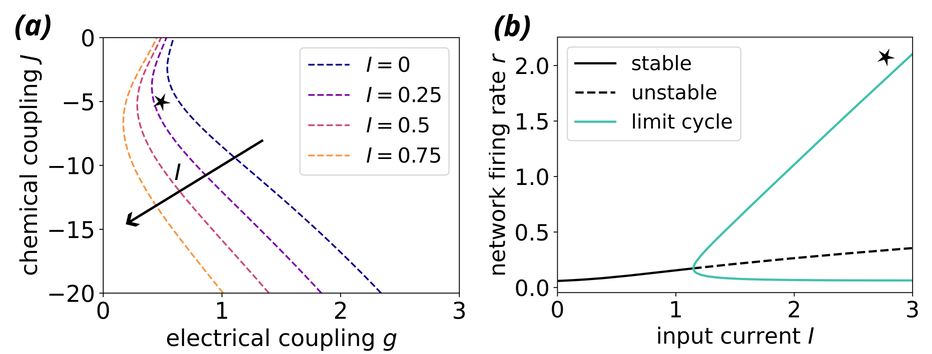}
\caption{Increased mean homogeneous input $I$ advances oscillations and increases the firing rate oscillatory amplitude. \textbf{(a)} Bifurcation diagram of chemical $J$ versus electrical $g$ coupling strengths for different values of input $I$ with $\bar{\eta}$ set to 1. Dashed lines represent the Hopf bifurcation. \textbf{(b)} Bifurcation diagram showing how the input $I$ affects the network firing rate. Parameters: $J=-5, g=0.5, \bar{\eta}=0, \Delta=0.3.$}
\label{fig:bif_I1} 
\end{figure}

\begin{figure}[H]
\centering
\includegraphics[width=\linewidth]{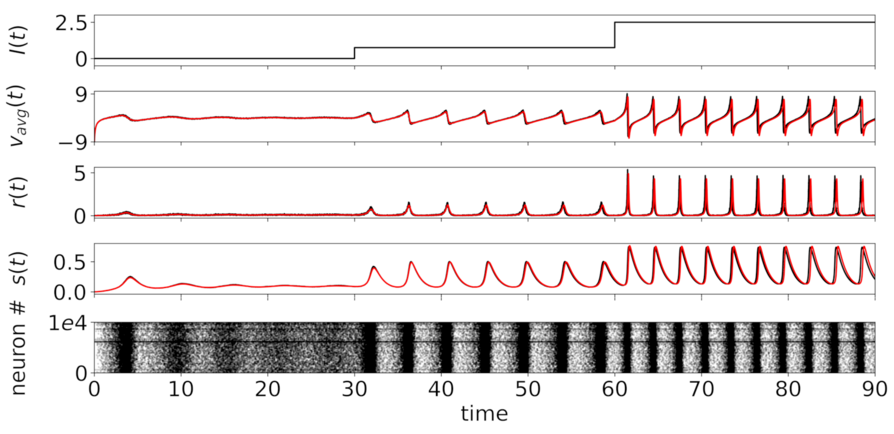}
\caption{Increasing input $I$ causes the network to transition from a stable focus regime to a limit cycle regime. Mean field equations (red) were plotted against numerical simulations (black). Parameters: $J=-10, g=1, \bar{\eta}=1, \Delta=0.3.$}
\label{fig:bif_I2}
\end{figure}

Increasing the input $I$ reduces the steady-state regime and thus expands the range of values $g$ and $J$ over which the system is in the stable limit-cycle regime (Fig. \ref{fig:bif_I1}a). Furthermore, as can be expected, for fixed values of $g$ and $J$, increasing the input current $I$ increases the amplitude of the network firing rate oscillations (Fig. \ref{fig:bif_I1}b). Taken together, these results demonstrate that increasing input current (e.g., in steps; Fig. \ref{fig:bif_I2}), drives the system rapidly from asynchrony to synchronous oscillations. We note that progressively larger currents simultaneously increase the amplitude and the frequency of population oscillations, thus providing a potential mechanism for the emergence of synchronisation within a single population.

\subsection{Increased heterogeneity delays oscillations} \label{sub:hetero}

\begin{figure}[H]
\centering
\includegraphics[width=\linewidth]{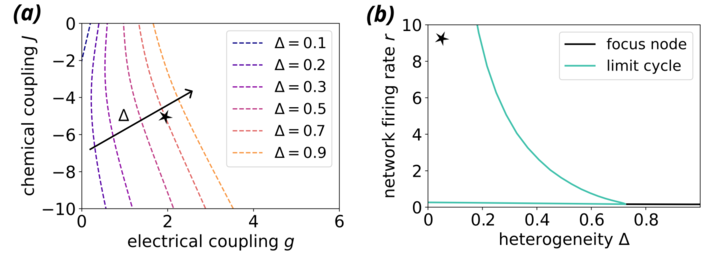}
\caption{Increased heterogeneity delays oscillations. \textbf{(a)} Bifurcation diagram of chemical $J$ versus electrical $g$ coupling strengths for different values of heterogeneity $\Delta.$ Dashed lines represent the Hopf bifurcation. \textbf{(b)} Bifurcation diagram showing how the heterogeneity $\Delta$ affects the network firing rate for fixed coupling strengths $g$ and $J$. Parameters: $J=-5, g=2, \bar{\eta}=1.$}
\label{fig:bif_delta}
\end{figure}

In biological networks, neurons present non-identical intrinsic properties that can be captured in this model through the distribution of heterogeneous input currents $\eta_j$, i.e. frozen noise, with mean absolute deviation $\Delta$. Network heterogeneity is therefore captured by $\Delta$, as larger values of $\Delta$ lead to neurons with greater variance in their excitability. We find that as the neural heterogeneity $\Delta$ increases, so does the range of values $g$ and $J$ such that the system is in a stable focus regime, as the bifurcation line shifts towards the right (Fig. \ref{fig:bif_delta}a). Therefore, for given fixed chemical $J$ and electrical $g$ coupling strengths, increasing the heterogeneity $\Delta$ shifts the bifurcation point towards stronger electrical coupling. We further note that oscillations cannot occur without the presence of electrical coupling except in conditions of low heterogeneity, in which case the network behaves as a single neuron receiving constant positive input current. Therefore, electrical coupling plays a key role in making oscillations robust to heterogeneous inputs. 

The heterogeneity also has an effect on modulating the network firing rate of the system when in the limit cycle regime. Indeed, we observe that as $\Delta$ decreases, the network firing rate $r$ increases (Fig. \ref{fig:bif_delta}b); this result is to be expected in light of previous work \cite{kopell2004, montbrio2019}, as decreasing the heterogeneity makes the neurons more alike and thus more likely to fire synchronously when injected with positive inputs. In what follows, we therefore consider a sufficiently large value of heterogeneity $\Delta > 0.2$ to account for the considerable heterogeneity found in biological neural networks.

\subsection{Summary of network properties impact on synchrony}
The effects of network properties (i.e.,  coupling strengths, time constants, input currents and heterogeneity) on network synchronisation is summarised in the following table:
\begin{table}[H]
\centering
\begin{tabular}{ll}
\hline
\multicolumn{1}{|l|}{\textbf{Parameter}}                                                       & \multicolumn{1}{l|}{\textbf{Effect on synchronisation}} \\ \hline
\multicolumn{1}{|l|}{\begin{tabular}[c]{@{}l@{}}Chemical coupling $J$\end{tabular}} & \multicolumn{1}{l|}{\begin{tabular}[c]{@{}l@{}} Increasing inhibitory chemical coupling \\ decreases oscillation frequency. \end{tabular}} \\ \hline
\multicolumn{1}{|l|}{\begin{tabular}[c]{@{}l@{}}Electrical coupling $g$\end{tabular}} & \multicolumn{1}{l|}{\begin{tabular}[c]{@{}l@{}} High enough electrical coupling favours \\ synchronisation. \end{tabular}} \\ \hline
\multicolumn{1}{|l|}{\begin{tabular}[c]{@{}l@{}}Asymmetry $a$\end{tabular}} & \multicolumn{1}{l|}{\begin{tabular}[c]{@{}l@{}}Increasing $a$ advances oscillatory behaviour.\end{tabular}}           \\ \hline
\multicolumn{1}{|l|}{\begin{tabular}[c]{@{}l@{}}Synaptic activation \\ time constant $\tau_d$\end{tabular}} & \multicolumn{1}{l|}{\begin{tabular}[c]{@{}l@{}}Increasing $\tau_d$ hinders synchronisation.\end{tabular}}           \\ \hline
\multicolumn{1}{|l|}{\begin{tabular}[c]{@{}l@{}}Homogeneous and \\ heterogeneous input \\ currents $I$ and $\bar{\eta}$\end{tabular}} & \multicolumn{1}{l|}{\begin{tabular}[c]{@{}l@{}}Increasing either $\bar{\eta}$ or $I$ advances \\ oscillations.\end{tabular}}  \\ \hline
\multicolumn{1}{|l|}{\begin{tabular}[c]{@{}l@{}}Mean heterogeneity $\Delta$\\ within the network \end{tabular}} & \multicolumn{1}{l|}{\begin{tabular}[c]{@{}l@{}}Increasing $\Delta$ delays oscillations.\end{tabular}} \\ \hline &
\end{tabular}
\end{table}

\section{INTERCOUPLED CLUSTERS} \label{sec:3}

Motivated by experimental data showing that electrical coupling between molecular layer interneurons may be organized in spatially localized cliques \cite{kim2021} with longer range inhibitory synaptic connections, we next consider the dynamics of interconnected clusters of neurons. In particular, we asked under what inter-cluster connectivity conditions would we expect the gap-junction coupling to lead to local synchrony with a single cluster versus global synchrony across clusters. As a minimal system, we consider two symmetrical clusters, with each cluster made up of all-to-all coupled neurons with electrical coupling $g_k$ and chemical coupling $J_k$ for $k \in \{1,2\}$. Furthermore, each cluster is intercoupled to the other through inhibitory chemical coupling $J_c$ (Fig. \ref{fig:sym_pop}a). We now examine the behaviours that occur when the two clusters are connected through such inhibitory synapses. In particular, we explore the emergence of network synchronisation and desynchronisation under transient input currents within this configuration.

\subsection{Network equations}
We consider interconnected clusters, with each individual cluster composed of $N_k$ neurons that are all-to-all connected by chemical synapses of coupling strength $J_k$ and gap junctions of coupling strength $g_k.$ The clusters are themselves interconnected via inhibitory synapses of coupling strength $J_c.$

\noindent Let $j \in \{1,..., N_k\}$ denote the index of neuron $j$ in cluster $k \in \{1,2\}$. The microscopic equations are given by the ODEs:

\begin{equation} \label{eq:clusters_micro}
\begin{split}
\tau_m \dot{v}_{j,k}(t) =& \ v_{j,k}(t)^2 + J_k \tau_m s_k(t) + g_k(\bar{v}_k-v_{j,k}(t)) \\ & + J_c \tau_m s_{3-k}(t) + \eta_{j,k} + I_k(t), \\[.5em]
\tau_d \dot{s}_k(t) =& \ -s_k(t) + r_k(t).
\end{split}
\end{equation}

\noindent Using the same exact reduction methods as in Appendix V A, we derive the macroscopic equations for cluster $k \in \{1,2\}$ as: 

\begin{equation} \label{eq:clusters_mf}
\begin{split}
\tau_m \dot{r}_k(t) =& \ \dfrac{\Delta}{\tau_m \pi} + 2r_k(t)v_k(t) - 2\tau_m \ln(a) r_k(t)^2 \\
& - g_k r_{k}(t), \\[.5em]
\tau_m \dot{v}_k(t) =& \ v_k(t)^2 + \bar{\eta} + J_k\tau_m s_k(t) + I_k(t) + J_c\tau_m s_{3-k}(t) \\
& + \dfrac{\Delta\ln(a)}{\pi} - (\ln(a)^2 + \pi^2)(\tau_m r_k(t))^2, \\[.5em]
\tau_d \dot{s}_k(t) =& \ -s_k(t) + r_k(t).
\end{split}
\end{equation}

\subsection{Entrainment of synchrony in clustered networks} 

\begin{figure}[H]
\centering
\includegraphics[width=.9\linewidth]{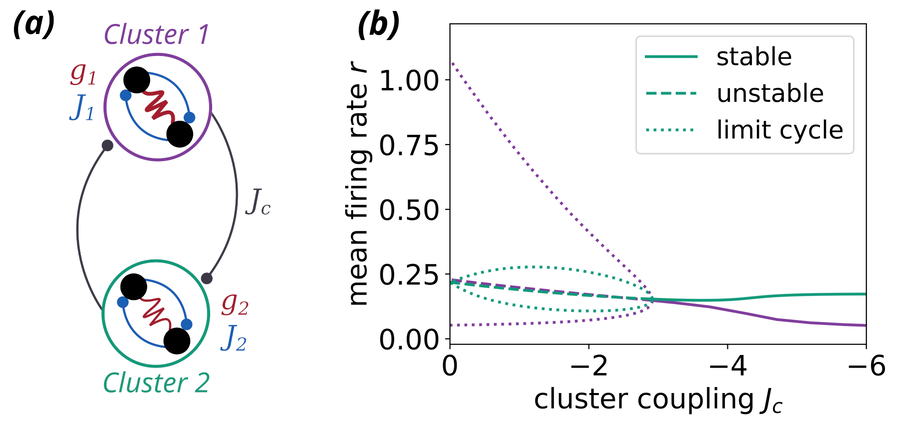}
\caption{One oscillating cluster triggers the other to oscillate through inhibitory synaptic coupling. \textbf{(a)} Illustration of the cluster connectivity. \textbf{(b)} Bifurcation diagram of network firing rate $r$ as a function of coupling strength $J_c$ between the two clusters (purple: cluster 1, green: cluster 2). Parameters: $g_1=1, g_2=0.4, J_1=J_2=-2.5, \bar{\eta}=1, \Delta=0.3.$}
\label{fig:sym_pop}
\end{figure}

We next asked under what conditions one cluster could entrain another cluster into synchronous oscillations. To do so, we begin by setting each cluster in one of each regimes from Fig. \ref{fig:bifs} in order to examine the effect of inhibitory synaptic coupling between both clusters. Cluster 1 is set in the Syn (i.e. limit cycle) regime, and cluster 2 in the Asyn (i.e. stable focus) regime. Without the presence of synaptic coupling between the two clusters ($J_c=0$), each cluster behaves as an individual population of QIF neurons, as analysed previously. Setting the coupling between the two populations at an intermediate range (Fig. \ref{fig:sym_pop}b; $J_c \in (-3,0)$) induces cluster 2 to begin synchronising at the same global frequency as cluster 1 (Supplementary S10). For stronger inhibitory coupling values (Fig. \ref{fig:sym_pop}b; $J_c < -3$) between the two clusters, the oscillations fade out in both clusters, resulting in asynchronous firing. 

\subsection{Interaction dynamics of cell-type specific clusters} 

\begin{figure}[H]
\centering
\includegraphics[width=\linewidth]{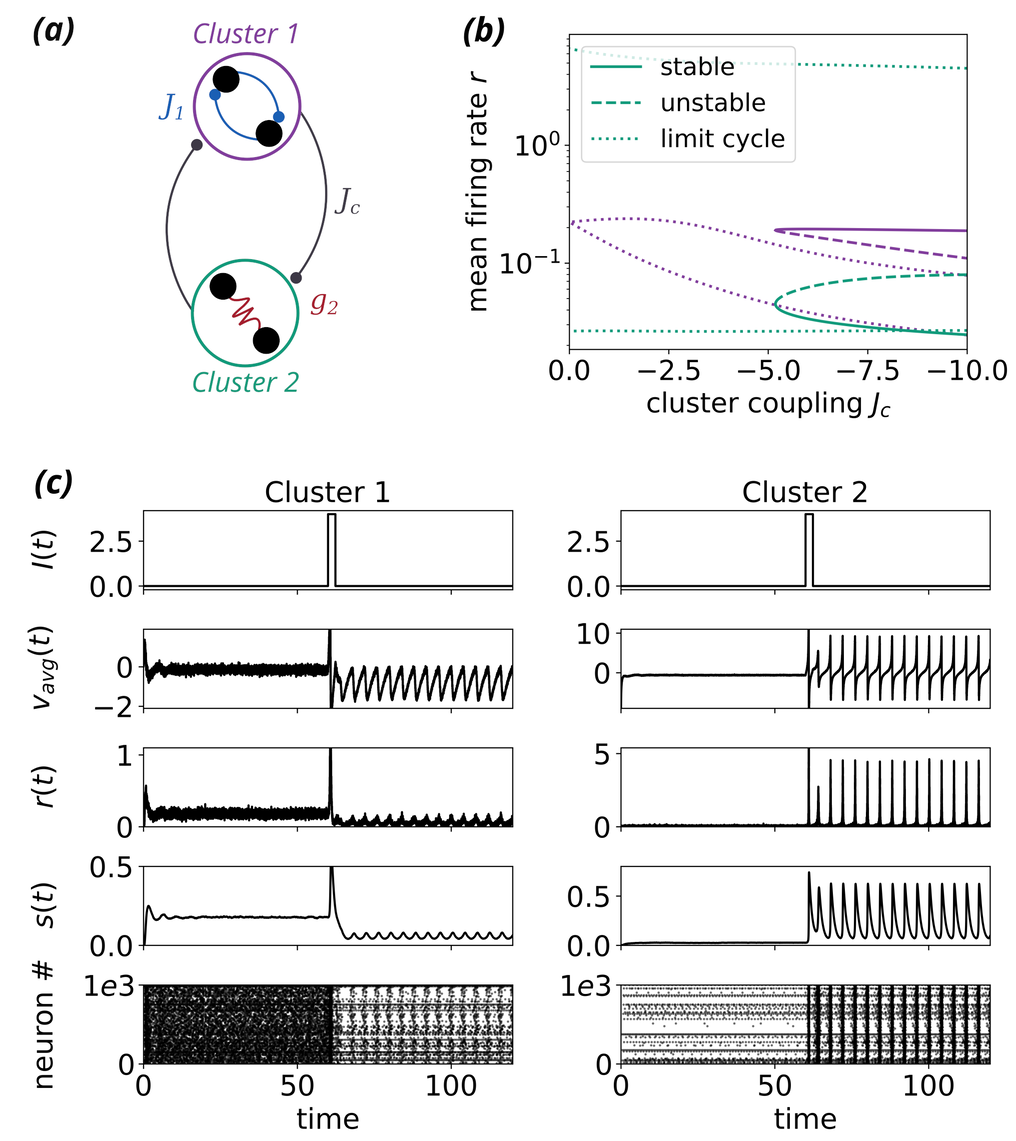}
\caption{Purely chemical and electrical clusters induce bistability. \textbf{(a)} Illustration of the cluster connectivity. \textbf{(b)} Bifurcation diagram as a function of coupling strength $J_c$ between the clusters. When $J_c< J_c^* \approx -5$, bistability emerges. \textbf{(c)} Network simulation illustrating the bistability which occurs when the coupling strength between the clusters is set to $J_c=-8$. Here, a pulse-like input switches the clusters from a stable asynchronous to a stable synchronous regime. Parameters: $g_1=0, g_2=2, J_1=-2.5, J_2=0, \bar{\eta}=1, \Delta=0.3.$ }
\label{fig:pop_bistability} 
\end{figure}

Paired recordings suggest that different types of cerebellar molecular layer interneurons form subnetworks where different coupling types predominate \cite{kim2021, kozareva2021}. Molecular layer interneurons are typically categorized into two distinct cell types: basket cells, which are predominantly interconnected via gap junctions, and stellate cells, which are interconnected via inhibitory chemical synapses \cite{alcami2013, hoehne2020}. Furthermore, these two groups are intercoupled with each other via inhibitory chemical synapses. Motivated by these biological properties we therefore analyse how two such clusters behave in the neural mass model framework (Fig. \ref{fig:pop_bistability}a). 

We find that for networks where each cluster is dominated by chemical connectivity or electrical connectivity, bistability emerges once the strength of inhibitory coupling between the two clusters reaches a critical value (Fig. \ref{fig:pop_bistability}b; $J_c^*$). Indeed, depending on the initial conditions, the clusters will either show a steady state behaviour (asynchronous firing in the full spiking network), or both clusters will fire synchronously together. Using a brief pulse current injected into both clusters, we can therefore switch between both behaviours (Fig. \ref{fig:pop_bistability}c). 

\begin{figure}[H] 
    \centering
    \includegraphics[width=\linewidth]{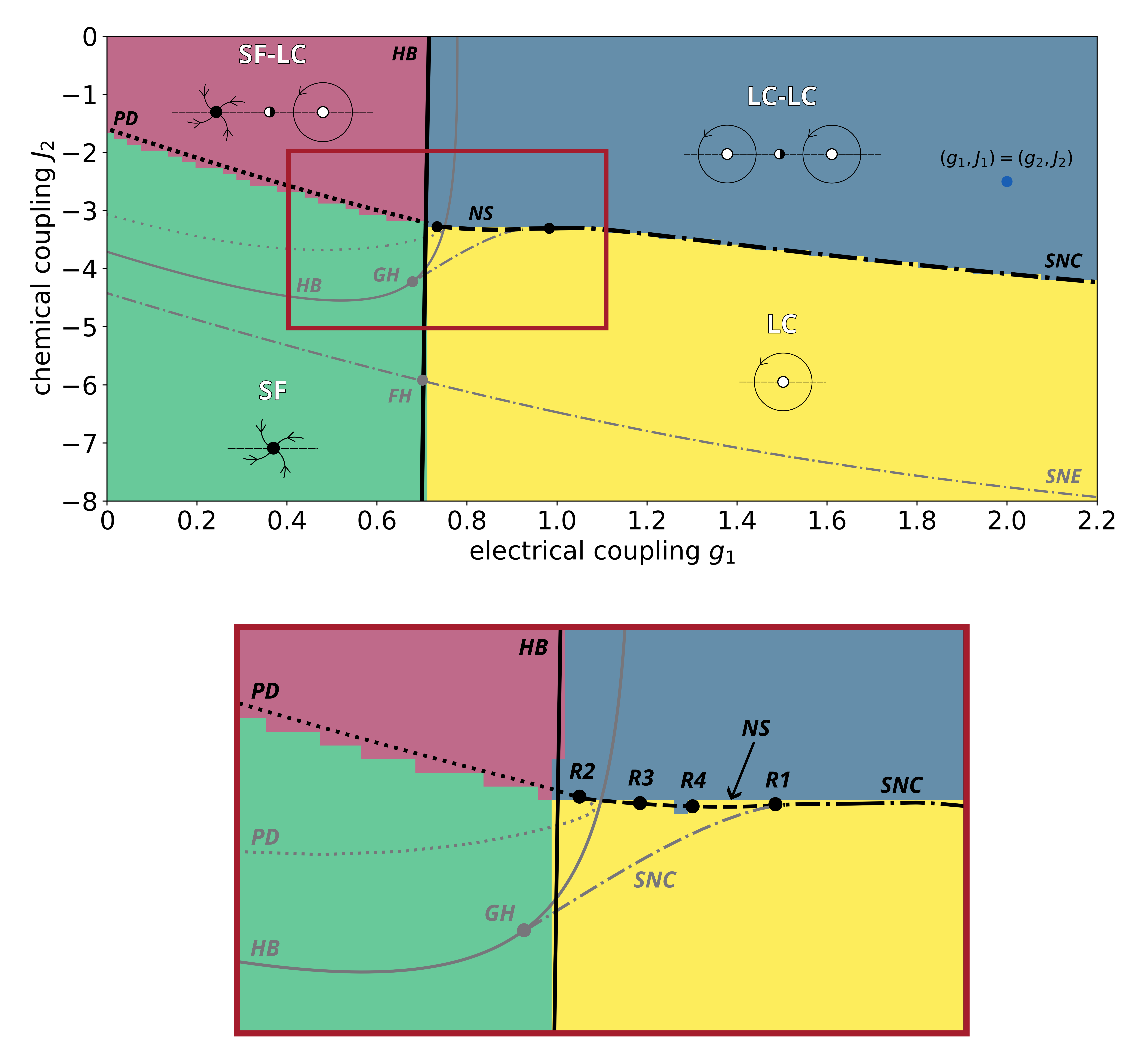}
    \caption{State diagram (coloured areas) and bifurcation curves describing the system's behaviour depending on cluster 1 electrical coupling $g_1$ and cluster 2 chemical coupling $J_2$ parameters. States consist of two bistable regimes: SF-LC (red) and LC-LC (blue), and two monostable regimes: SF (green) and LC (yellow). Bifurcations consist of supercritical period‐doubling (PD), supercritical Hopf (HB), saddle‐node of cycles (SNC), saddle‐node of equilibria (SNE) and Neimark Sacker (NS) with codimension-two bifurcation points consisting of generalized Hopf (GH) where SNC disappears by colliding with HB, 1:1 resonance (R1) on the SNC and 1:2 resonance (R2) on the PD where NS disappears by colliding with SNC and PD, and finally Fold-Hopf (FH) where SNE and HB intersect. Bold black lines indicate relevant bifurcations responsible for a change in stability of the attractors, coinciding with the change of observable states. Blue dot represents parameters for which both clusters are symmetrical. Parameters: $g_2=2, J_1=-2.5, J_c=-8, \bar{\eta}=1, \Delta=0.3$. } \label{fig:bif_cluster}
\end{figure}

To test the robustness of this bistability, we next explored what behaviours emerge when setting the system in the bistable regime (e.g., fixing the intercoupling strength $J_c=-8 < J_c^*$) and introduced electrical coupling $g_1$ in cluster $1$ as well as chemical coupling $J_2$ in cluster $2$. In order to assess the stable states in our system, we ran multiple simulations using the reduced mean-field equations (Eq. \ref{eq:clusters_mf}) and determined the nature of the steady-state solutions (see Supplementary V B). Interestingly, we found the existence of four different regimes (Fig. \ref{fig:bif_cluster}), depending on the coupling coefficients within each of the clusters: a bistable regime in which stable focus and limit cycle coexist (SF-LC), i.e., both clusters are either asynchronous or oscillate at same frequency; an oscillatory bistable regime in which two distinct limit cycles coexist (LC-LC), i.e., both clusters oscillate with same frequency and can switch between two distinct limit cycles; a steady state regime in which both clusters exhibit a stable focus (SF) and a limit cycle regime in which both clusters oscillate (LC) (see Supplementary S14 for numerical simulations within each regime).

In order to better understand the dynamics leading to these different regimes, we analyse the bifurcations occurring (using AUTO). We plotted the bifurcation curves over the state diagram previously obtained (Fig. \ref{fig:bif_cluster}) and found that these bifurcations were able to explain the transitions between the four different states. We begin by noticing that the vertical part of the Hopf bifurcation curve marks the transitions driven by increasing the electrical coupling: where a focus loses stability to a limit cycle, providing an explanation for the transition from the SF-LC to the LC-LC regime and the SF-LC transition. The other part of the Hopf bifurcation curve however represents a change in the saddle-focus equilibrium (see Supplementary S12) and is of no consequence in the change of stability of the steady-states.  For completeness we also note that the Hopf bifurcation extends beyond the range of values $J_2$ considered here, as it forms a loop in the area where $J_2 >0$ (see S11 for the extended bifurcation diagram). 

If we start in the SF-LC regime and decrease the synaptic coupling within cluster 2 which is predominantly gap-junction coupled $J_2$, the limit cycle loses stability through a supercritical period-doubling (PD) bifurcation (dotted line). We found that this bifurcation leads to cascades and chaotic orbits (see S13), causing the LC attractor to lose stability and collapse onto the SF attractor. In the LC-LC regime, the loss of stability in one of the limit cycles as $J_2$ decreases can be explained through a saddle-node of cycles (SNC) bifurcation and a Neimark Sacker (NS) bifurcation giving rise to an invariant torus attractor. Indeed, we found a 1:2 resonance bifurcation point (R2) on the PD and a  1:1 resonance bifurcation point (R1) on the SNC, which both mark the extremities of the NS (see Fig. S11). We also detected a 1:3 and a 1:4 resonance bifurcation point on NS; the branches emerging from these various resonance bifurcations (1:1, 1:3 and 1:4) may also play a role in shaping the dynamics of the attractors occurring near these bifurcations, but a more rigorous analysis will be needed in future work to determine the subtle nature of these dynamics. Furthermore, as we impose that $r_1>0$ and $r_2>0$, there are either three distinct equilibria (two of which are separated by a saddle) or a single equilibrium, depending on the coupling parameters $g_1$ and $J_2$. We would like to point out that the saddle-node of equilibria (SNE) bifurcation, while not affecting the limit cycle behaviour, demarcates the number of equilibria in our system - three equilibria above the SNE and one below it. Together, these results display the wealth of dynamics that can occur when incorporating non-homogeneous coupling structures into the connectivity properties of the network. 

To summarize the four regimes qualitatively, we observe two bistable regions at weak inhibitory coupling $J_2$: one in which a steady state and an oscillation co-exist (Fig. \ref{fig:bif_cluster}, red region) and another in which two different oscillatory cycles co-exist (Fig. \ref{fig:bif_cluster}, blue region). These two states are separated from each other by a supercritical branch of the Hopf bifurcation. For stronger inhibitory coupling $J_2$, bistability collapses through either period-doubling cascades (Fig. \ref{fig:bif_cluster}, red to green) leading to a stable focus, or through a Neimark Sacker/ saddle node of cycles bifurcation curves, leading to a single stable oscillation (Fig. \ref{fig:bif_cluster}, blue to yellow).

The rich multi-stable dynamics of the system that we identified above allow for rapid switching of the states by external controls. Hence, motivated by the fact that molecular layer interneurons receive excitatory inputs from granule cells \cite{kim2021, traub2001, vervaeke2010}, we propose a mechanism that allows to switch from asynchrony to synchronous firing within the two clusters upon the brief injection of an input current to one cluster. Within the (SF-LC) regime, a brief positive external current to the gap junction cluster (i.e. cluster 2) allows for a switch in stability from asynchrony to synchronous firing (Supplementary S15b). This however is not true for the synaptic connected cluster (i.e. cluster 1) (Supplementary S15a). On the other hand, oscillations can only be shut down by applying this external current to the inhibitory synaptic cluster (Supplementary S16a). Note that in the case of a brief negative external current, the converse is true (Supplementary S15c and S16d). These results therefore provide a potential mechanism by  which transient excitatory activation can switch between synchronous and asynchronous regimes by targeting distinct cell-type specific clusters. \\

\section{DISCUSSION}
In this paper we investigated how the interplay of multiple coupling and intrinsic neuronal properties impact the presence of synchronous oscillations in interneuron networks. In particular, we were motivated by mechanisms for rapid input-dependent synchronisation of inhibitory interneurons, which are abundant across brain regions and which are often interconnected through both chemical synapses and gap junctions. Towards this end, we used an exact reduced mean-field approach combined with bifurcation analysis to identify parameter regimes supporting stable limit cycle dynamics. We first unified the previously obtained, at times disparate, results from the neural mass literature \cite{montbrio2015, montbrio2017, montbrio2019, montbrio2020} into a comprehensive conceptual framework of the neural mass model of an all-to-all connected, homogeneous population with combined chemical and electrical coupling, slow synapses and asymmetric reset values.

We find that generically in such populations, two possible stable behaviours arise: either the neurons fire asynchronously, leading to attractor dynamics in the population firing rate, or they synchronise globally, corresponding to a limit cycle. Furthermore, synchronisation is driven primarily by sufficiently strong electrical gap-junction coupling to transition the network through a supercritical Hopf bifurcation, by increasing either the gap-junction strength or the input current amplitude. Additional parameters linked to the intrinsic properties of the constituent neurons, such as heterogeneity, asymmetry of the spike, and time constants, affect the position of the Hopf bifurcation without changing the network's overall behaviour. The strong dependence of limit cycle dynamics on strong electrical coupling in our model is consistent with previous work highlighting the importance of gap junctions for synchrony. Interestingly, however, we also find that the frequency of synchronous oscillations depends solely on the strength of inhibitory coupling. These results suggest a potential role for the presence of chemical synapses in electrically coupled networks, as observed in diverse interneuron populations across the brain \cite{galarreta2001}. 

Our biological motivation to examine the mechanisms leading to synchrony in chemically and electrically coupled networks was inspired by the circuitry of cerebellar molecular layer interneurons (MLIs). Rapid, input-dependent synchronisation of MLIs could synchronise Purkinje cells, the output neuron of the cerebellar cortex, helping them to recruit downstream nuclear neurons \cite{person2012}. One possible dynamical mechanism for such rapid synchronisation in MLI networks would be a bistable network in which a fixed point attractor and a limit cycle coexist in the mean-field dynamics. In this case, a rapid, transient external input would be sufficient to transition the network into a synchronous regime. Interestingly, however, we did not observe bistability in these all-to-all coupled interneuron networks for any of the extensive parameter regimes here examined. Therefore, within the single-population configuration, transient input pulses cannot induce synchrony, and only a step current can induce synchronisation from an asynchronous state.

However, in the cerebellum, MLIs do not form a single homogeneous population but are rather separated into two morphologically and structurally defined cell types, basket cells and stellate cells, which form clustered networks that tend to be dominated by electrical coupling (between basket cells) or chemical coupling (between stellate cells). We therefore next considered networks with clustered connectivity corresponding to different cell types, which are themselves interconnected through inhibitory synapses. Specifically, we modelled the connectivity within each cluster such that one was dominated by gap junctions and the other by inhibitory synapses, in line with experimentally observed preferential electrical versus chemical coupling in basket and stellate cells. We find that for such networks several complex behaviours emerge. Firstly, we find that one cluster can induce another to oscillate with same phase given strong enough inhibitory synaptic coupling between the two. Secondly, we find bistability when one cluster is dominantly connected by electrical coupling and the other by chemical coupling, much like basket cells versus stellate cells. More precisely, we find four possible dynamical regimes depending on the connectivity strengths, that are common to both clusters: 1) a bistable regime in which a steady state and stable limit cycle coexist (i.e., both clusters are either asynchronous, or  both oscillating at phase-locked frequency), 2) an oscillatory bistable regime in which two distinct limit cycles coexist, 3) a steady state regime in which both clusters are in a steady state, and 4) a limit cycle regime in which both clusters oscillate. In the bistable regimes, a pulse current projected to the network as a whole can allow for a rapid switch between asynchrony and synchrony within the two clusters. 

Such a bistable regime therefore provides a potential substrate through which an external input can rapidly transition between synchronous and asynchronous regimes in clustered interneuron networks. Interestingly, in the cerebellum, stellate cells and basket cells both receive feedforard excitation from parallel fibres inputs \cite{cayco-gajic2019}. This feedforward excitation, as observed in our model, allows for a rapid switch between synchrony and asynchrony of the network upon receiving a brief positive input current. Moreover, due to their spatial organisation (stellate cells are found mostly in the upper molecular layer whereas basket cells are found mostly in the lower molecular layer), basket cells also receive sparse feedback inhibition from the collaterals of adjacent Purkinje cells. By introducing this organisation of clustered connectivity in our model (i.e. one cluster predominantly exhibiting chemical synapses and the other predominantly exhibiting gap junctions) demonstrates that a brief \emph{inhibitory} input current targeting the basket cell cluster (predominantly gap-junctional connections) can cause suppression of synchronous spiking in both clusters. Interestingly, this result mirrors experimental findings both in vitro and in vivo recordings of the mouse cerebellum \cite{halverson2022}. Based on our results, we speculate a possible mechanism by which excitatory input from parallel fibres could cause rapid-onset synchrony in MLIs by transitioning average network activity from the stable fixed point to the limit cycle, after which inhibition from Purkinje cell collaterals onto basket cells could rapidly return the network back to the asynchronous steady state. This could provide a mechanism for rapid, transient synchrony in MLI networks that could help recruit downstream populations. 

While we have focused on developing a neural mass model to consider clustered cerebellar MLI networks, future work could extend upon this framework, especially in light of more recent contributions to the exact mean-field reduction. These include the addition of sparse connectivity \cite{divolo2020}, noisy input current \cite{montbrio2024}, short-term adaptation \cite{gast2020, chen2022}, as well as intercoupling between many populations \cite{gerster2021}, spatio-temporal dynamics \cite{coombes2021} and an exact mean-field reduction of the Izhikevich QIF model \cite{izhikevich2003} which could offer insight into the impact of a richer variety of intrinsic neuronal dynamics \cite{guerreiro2023, gast2024}. Together, these new advances conspire to pave an exciting path forward for exact mean-field models that merge analytical tractability with increasing biological realism. \\

\section*{Acknowledgements}
This research was funded by Agence Nationale pour la Recherche (ANR-17-EURE-0017, ANR-10IDEX-0001-02, InTemp), ENS, CNRS and INSERM.\\

MD is supported by the grant PID2023-146683OB-100 funded by MICIU/AEI /10.13039/501100011033 and by ERDF, EU. Additionally, MD is supported by the Basque Government through the BERC 2022-2025 program and by the Ministry of Science and Innovation: BCAM Severo Ochoa accreditation CEX2021-001142-S / MICIU / AEI / 10.13039/501100011033. Moreover, MD acknowledges support of SILICON BURMUIN no. KK-2023/00090 funded by the Basque Government through ELKARTEK Programme.

\bibliography{biblio.bib} 
\clearpage
\onecolumngrid

\section{APPENDIX}
\setlength\parindent{0pt}

\subsection{From Microscopic to Macroscopic} \label{proof:mean_field}
For the reader, we present here a unified proof of the exact mean-field reductions \cite{montbrio2015} that has been individually derived for gap junctions \cite{montbrio2019}, slow chemical synapses \cite{montbrio2017} and asymmetric resets \cite{montbrio2020}. We have that $\rho(v|\eta, t)dv$ is the fraction of neurons with membrane potentials between $v$ and $v+dv$ with parameter $\eta$ at time $t$. Therefore, $\int_{-\infty}^{+\infty}\rho(v|\eta,t)c(\eta)d\eta$ is the total voltage density at time $t.$ In order for there to be conservation of all neurons in the system, $\rho(v|\eta,t)$ must verify the continuity equation: 
\begin{equation} \label{eq:CE}
	\tau_m \partial_t \rho + \partial_v [ (v^2 + \eta + g(\bar{v}-v) + J\tau_m s + I) \rho ] = 0.
\end{equation} 

Supposing the following ansatz (later referred to as Lorentzian ansatz or LA) - the solution to (\ref{eq:CE}) is of the form:
\begin{equation}
	\label{eq:ansatz}
	\rho(v|\eta,t) = \frac{1}{\pi}\frac{x(\eta, t)}{(v-y(\eta, t))^2 + x(\eta,t)^2} \cdot
\end{equation} 

The firing rate $r(t)$ is given by the probability flux evaluated at threshold voltage, i.e. $r(\eta, t) = \rho(v \to \infty | \eta, t)\dot{v}(v \to \infty |\eta,t)$: 
\begin{align}
	r(\eta,t) &= \lim_{v \to \infty}\rho(v|\eta, t)\dot{v}(v|\eta,t) = \lim_{v \to \infty} \frac{1}{\pi}\frac{x(\eta, t)(v^2 + \eta + g(\bar{v}-v) + J\tau_m s + I)}{(v-y(\eta,t))^2+x(\eta,t)^2\tau_m} = \frac{x(\eta,t)}{\tau_m\pi},
\end{align}
and it immediately follows that: 
\begin{equation} \label{eq:r}
	r(t) = \int_{-\infty}^{+\infty} r(\eta,t)c(\eta)d\eta =  \int_{-\infty}^{+\infty} \frac{x(\eta,t)}{\tau_m\pi}c(\eta)d\eta.
\end{equation}

By definition of the LA as a distribution of mean $y(\eta,t),$ the mean membrane potential w.r.t. $\eta$ is: 
\begin{equation}
	y(\eta,t) = p.v. \int_{-\infty}^{+\infty} \rho(v|\eta,t)vdv = \lim_{R \to +\infty} \int_{-R}^{R} \rho(v|\eta,t)vdv,
\end{equation}
and by definition of the expectancy of the voltage for neurons with a certain $\eta$ value: 

\begin{align} 
	\ v(\eta, t) =  \lim_{v_r \to +\infty} \int_{-v_r}^{v_{th}=av_r} \rho(v|\eta,t) v dv = \underbrace{p.v. \int_{-\infty}^{+\infty} \rho(v|\eta,t)vdv}_{= y(\eta,t)} + \underbrace{\lim_{v_r \to +\infty} \int_{vr}^{av_r} \rho(v|\eta,t)vdv}_{\coloneqq (\star)}.
\end{align}

Let us compute $(\star)$:

\begin{align}
	\int_{vr}^{av_r} \rho(v|\eta,t)vdv =& \int_{v_r}^{av_r} \frac{x(\eta,t)}{(v-y(\eta,t))^2+x(\eta,t)^2}vdv = \ x(\eta,t) \int_{v_r-y}^{avr-y} \frac{u+y}{u^2+x^2}du \text{, with $u=v-y$,  }\\[0.75em]
	=& \ x(\eta,t) \left[ \frac{\ln(u^2+x^2)}{2} \right]_{u=v_r-y}^{u=av_r-y} + x(\eta,t) \left[ \frac{y \arctan(t)}{x} \right]_{t=\frac{v_r-y}{x}}^{t=\frac{av_r-y}{x}} \cdot
\end{align}

Therefore, as $v_r \to \infty$, we obtain the following:  
\begin{equation}
	v(\eta,t) = y(\eta,t) + \frac{\ln(a)}{\pi}x(\eta,t),
\end{equation}

and it immediately follows that: 
\begin{align}
	v(t) = \int_{-\infty}^{+\infty} v(\eta,t)c(\eta)d\eta = \int_{-\infty}^{+\infty}y(\eta,t)c(\eta)d\eta + \tau_m\ln(a)r(t). \label{eq:v}
\end{align} 

Injecting (\ref{eq:ansatz}) into (\ref{eq:CE}), we obtain: 
\begin{align} 
	& v^2(\tau_m\dot{x} - 2xy + gx) + v(-2\tau_m\dot{x}y + 2\tau_m x\dot{y} + 2xy^2 + 2x^3 - 2x(\eta+J\tau_m s+I+g\bar{v}) ) + \tau_m\dot{x}y^2 - \tau_m x\dot{x} - 2\tau_m x\dot{y}y - gxy^2 \nonumber \\
	& - gx^3 + 2xy(\eta+J\tau_m s +I+g\bar{v}) = 0.
\end{align}

For this equality to be true regardless of the value of $v$, we require that each term be equal to $0.$ Therefore, we end up with the following system: 
\begin{align}
	\tau_m \dot{x} &= 2xy - xg, \\
	\tau_m \dot{y} &= y^2 - x^2 + \eta + J\tau_m s + I + g(\bar{v}-y).
\end{align}

The system can be solved by rewriting it in complex form, yielding: 
\begin{equation} \label{eq:mf_complex_form}
	\tau_m (\dot{x} + i \dot{y}) = \ i(\eta + J\tau_m s +I+g\bar{v}-(x+iy)^2) -g(x+iy).
\end{equation}

We now solve equations (\ref{eq:r}) and (\ref{eq:v}). Let us begin with equation (\ref{eq:r}): 
\begin{align}
	r(t) = \frac{1}{\tau_m\pi} \int_{-\infty}^{+\infty} x(\eta,t)c(\eta)d\eta = \frac{1}{\tau_m\pi} \int_{-\infty}^{+\infty} x(\eta,t)\frac{1}{\pi} \frac{\Delta}{(\eta-\bar{\eta})^2+\Delta^2}d\eta = \frac{\Delta}{\tau_m\pi^2} \int_{-\infty}^{+\infty} x(\eta,t)\frac{1}{(\eta-\bar{\eta})^2+\Delta^2}d\eta.
\end{align} 

Partial fraction decomposition yields: 
\begin{equation}
	\frac{1}{(\eta-\bar{\eta})^2 + \Delta^2} = \frac{1}{2i\Delta}\frac{1}{(\eta-\bar{\eta}-i\Delta)} - \frac{1}{2i\Delta}\frac{1}{(\eta-\bar{\eta}+i\Delta)} \cdot
\end{equation}

These fractions admit two poles $\eta_1 \coloneqq \bar{\eta}-i\Delta$ and $\eta_2 \coloneqq \bar{\eta}+i\Delta$ of order $|-1|$. We will use the notation $z_j = \eta_j, j \in \{1,2\}$ to denote the poles of our function. \\

To simplify notations, we will denote:
\begin{equation*} 
	h(\eta) \coloneqq x(\eta,t) \frac{1}{(\eta-\bar{\eta})^2 + \Delta^2} \cdot
\end{equation*}

Therefore, we integrate around the pole $\eta_1$ using Cauchy's residue theorem :
\begin{align}
	r(t) &= -\frac{\Delta}{\tau_m\pi^2} \ 2i\pi \sum_{\Im(z_j)<0} \text{Res}(h,z_j) = -\frac{2i\Delta}{\tau_m\pi} \frac{1}{(1-1)!} \lim_{\eta \to \eta_{1}} \frac{\partial^{1-1}}{\partial\eta^{1-1}}((\eta-\eta_{1})^1 h(\eta)) \\[.3em]
	&= - \frac{2i\Delta}{\tau_m\pi} \lim_{\eta \to \eta_{1}} (\eta-\eta_{1}) x(\eta,t) \frac{1}{(\eta-\bar{\eta})^2 + \Delta^2} = -\frac{2i\Delta}{\tau_m\pi} \lim_{\eta \to \eta_{1}} x(\eta,t) \left[ \frac{1}{2i\Delta}\frac{\eta-\bar{\eta}+i\Delta}{\eta-\bar{\eta}-i\Delta} - \frac{1}{2i\Delta} \right] \\[.4em]
	&= \frac{2i\Delta}{\tau_m\pi} \frac{1}{2i\Delta}x(\bar{\eta}-i\Delta,t) = \frac{1}{\tau_m\pi}x(\bar{\eta}-i\Delta,t).
\end{align}

Therefore, the network firing rate $r(t)$ is given by: 

\begin{equation} \label{eq:r_sol} 
	\centering{r(t) = \frac{1}{\tau_m\pi}x(\bar{\eta}-i\Delta,t). }
\end{equation}

We now move onto equation \ref{eq:v}. Using the same reasoning as above, the mean voltage $v(t)$ is given by: 
\begin{equation} \label{eq:v_sol}
	v(t) = y(\bar{\eta}-i\Delta, t) + \tau_m\ln(a)r(t). 
\end{equation} 
Injecting (\ref{eq:r_sol}) and (\ref{eq:v_sol}) into (\ref{eq:mf_complex_form}) and evaluating (\ref{eq:CE}) at $\eta = \bar{\eta}-i\Delta$, we now have the expression for the exact mean field reduction: 
\begin{align}
	& \tau_m (\tau_m \pi \dot{r}) = 2\tau_m\pi r(v-\tau_m\ln(a)r) - g\tau_m\pi r, \\[.5em]
	& i\tau_m (\dot{v}-\tau_m\ln(a)\dot{r})) = i(\bar{\eta} - i\Delta + J\tau_m s + I + g\bar{v} -\tau_m^2\pi^2r^2 + (v-\tau_m\ln(a)r)^2-g(v-\tau_m\ln(a)r)),
\end{align}
and end up with the following set of ODEs that describes the system: 
\begin{align} 
	\tau_m\dot{r}(t) =& \dfrac{\Delta}{\tau_m\pi} + 2r(t)v(t) - 2\tau_m\ln(a)r(t)^2 - gr(t), \\
	\tau_m\dot{v}(t) =& \ v(t)^2 + \bar{\eta} + J\tau_ms(t) + I - (\ln(a)^2 + \pi^2)(\tau_mr(t))^2 + \dfrac{\Delta\ln(a)}{\pi}, \\
	\tau_d\dot{s}(t) =& -s(t) + r(t). 
\end{align} 
\vspace{1em}

\subsection{Numerical methods} \label{supp:num_impl}

Numerical simulations were made using the Euler scheme with time step $\Delta t=0.001$. We fixed the reset value to $v_r=-100$ and $v_{th}$ was varied depending on the value of parameter $a$ considered. All the bifurcation diagrams presented in this paper were obtained using the bifurcation analysis softwares XPPAUT and AUTO on the mean-field equations. The state diagram in Fig. 9 was obtained by simulating the cluster mean-field equations (Eq. 5) with initial conditions close to equilibria and analysing the stability of the steady-state solutions. \\ 

All the code used to generate figures and results, as well as videos accompanying S\ref{fig:suppfig3} will be uploaded at \href{https://github.com/helene-todd/NMM2025}{https://github.com/helene-todd/NMM2025}.

\clearpage
\subsection{Supplementary Figures} \label{supp:figures}

\setcounter{figure}{9}  
\makeatletter
\renewcommand*{\fnum@figure}{{\normalfont\bfseries S\thefigure}}
\makeatother

\begin{figure}[H]
	\centering
	\includegraphics[width=.8\linewidth]{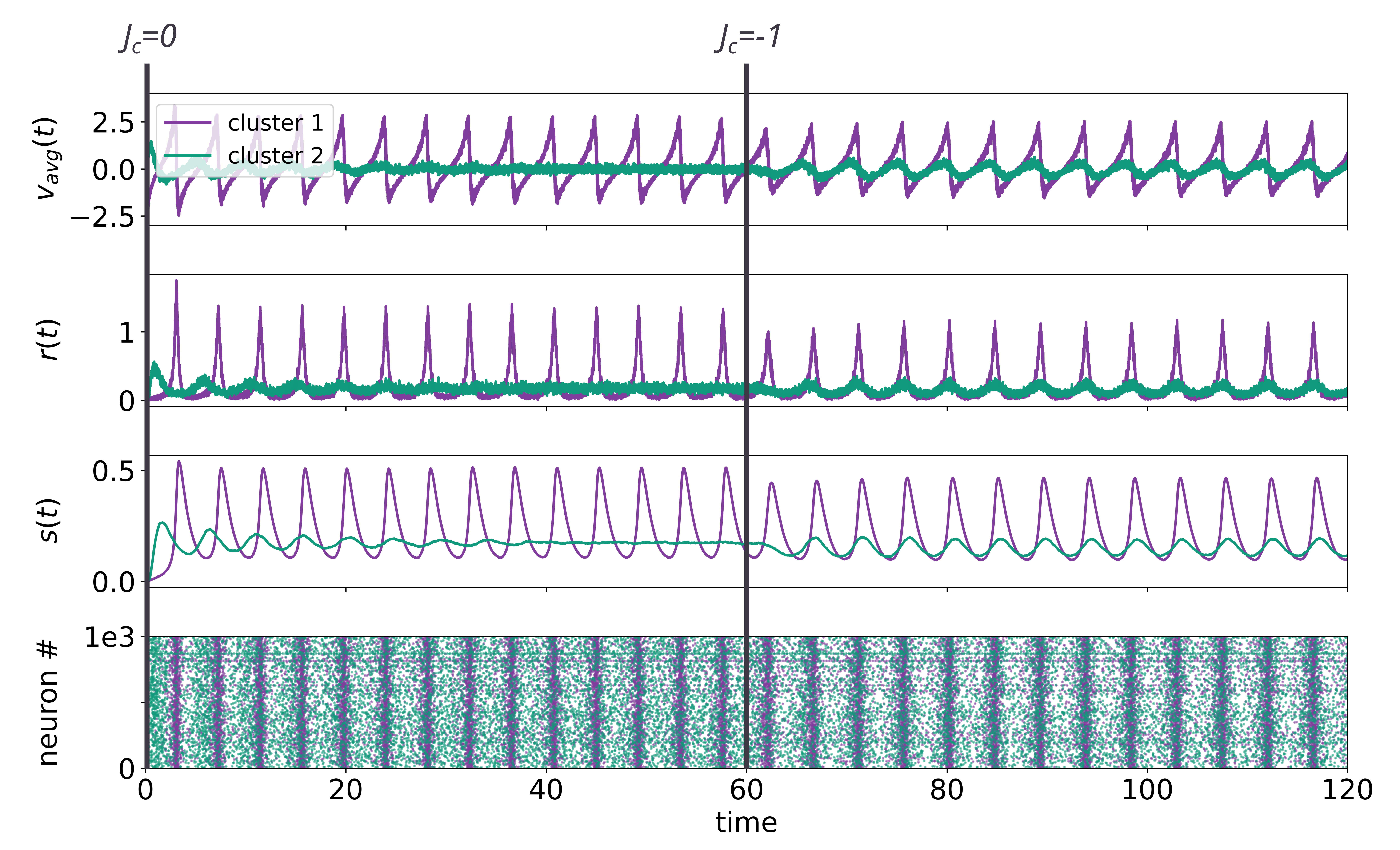}
	\captionsetup{width=.8\linewidth}
	\caption{One oscillating cluster induces the other to oscillate with same frequency as soon as they are coupled together through inhibitory synapses. Network simulation of $10^4$ neurons in each cluster, with nonzero cluster connectivity $J_c$ introduced at time $T = 60$. Cluster 1 starts in a LC regime while cluster 2 starts in a SF regime. Parameters: $g_1=1, g_2=0.4, J_1=-2.5, J_2=-4, J_c=0, \bar{\eta}=1, \Delta=0.3$.}
	\label{fig:suppfig1}
\end{figure}

\newpage
\begin{figure}[H]
	\centering
	\includegraphics[width=.65\linewidth]{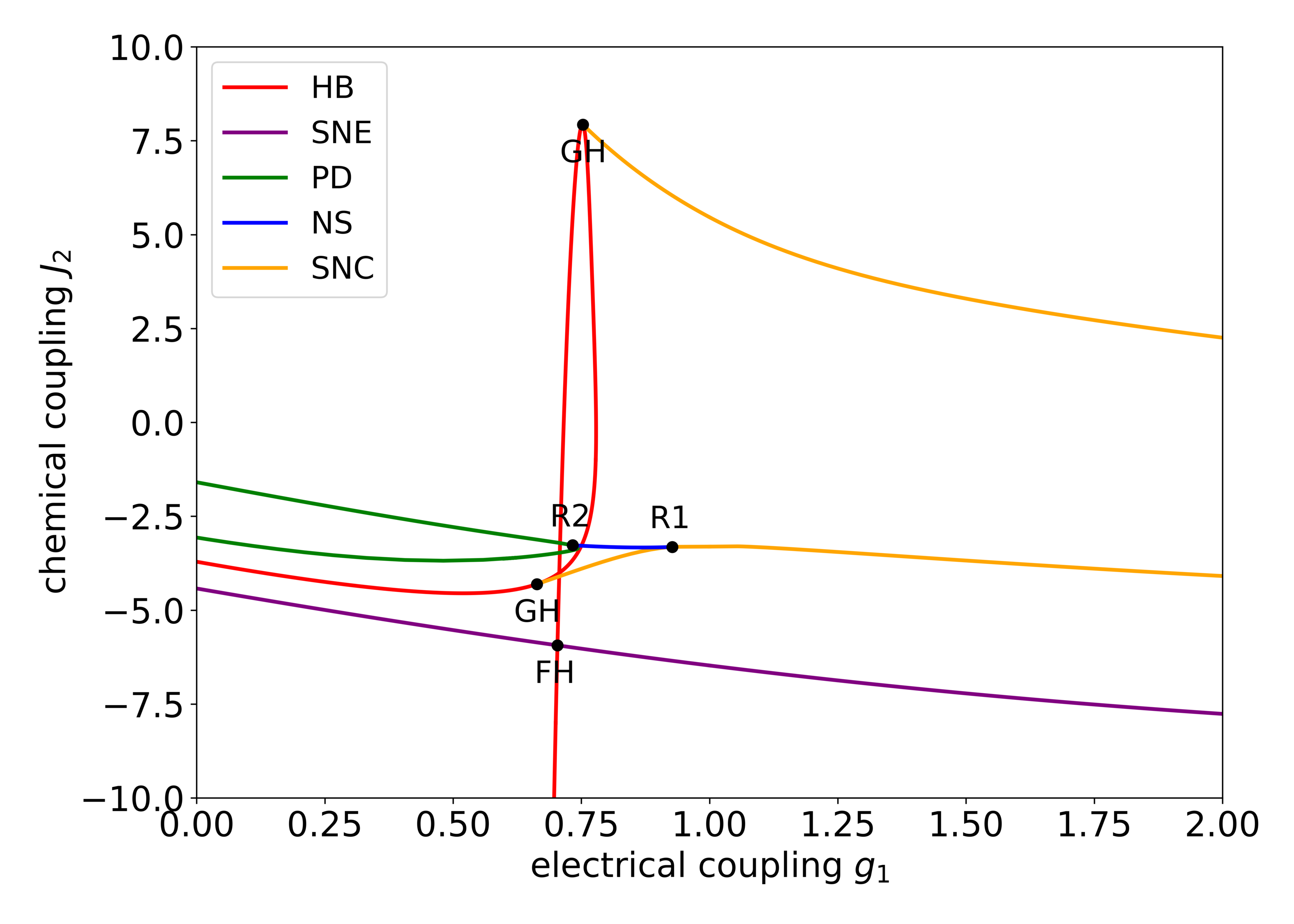}
	\captionsetup{width=.65\linewidth}
	\caption{Full bifurcation diagram beyond the scope of parameters $(g_1, J_2)$ considered in the main text. The HB forms a loop in the plane $J>0$, intersecting a SNC curve in a GH point at its peak.}
	\label{fig:suppfig2}
\end{figure}

\newpage
\begin{figure}[H]
	\centering
	\includegraphics[width=.6\linewidth]{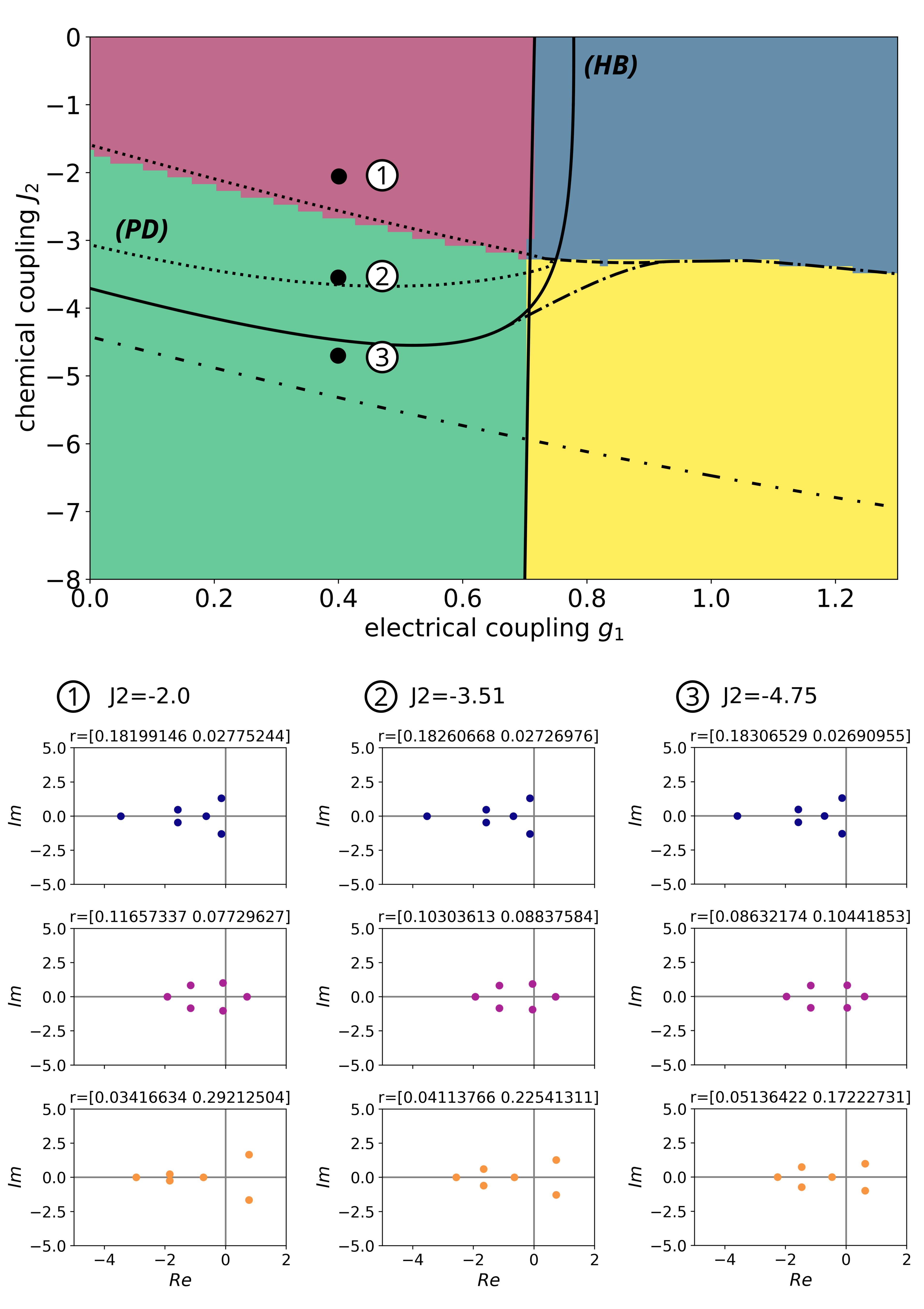}
	\captionsetup{width=.6\linewidth}
	\caption{Eigenvalues for various stable fixed points on the line $g_1=0.4.$ We observe that the outer loop of the (HB) reflects the complex conjugate eigenvalues of the saddle-focus (purple points) crossing the real axis. A similar analysis (not shown here, see Supplementary videos) was conducted on the $J_2=-2$ line while varying $g_1$ from $0$ to $1.3,$ in which we observe that the complex conjugate eigenvalues of the stable focus (blue points) cross the real line followed by the complex conjugate eigenvalues of the saddle-focus.}
	\label{fig:suppfig3}
\end{figure}

\newpage
\begin{figure}[H]
	\centering
	\includegraphics[width=.6\linewidth]{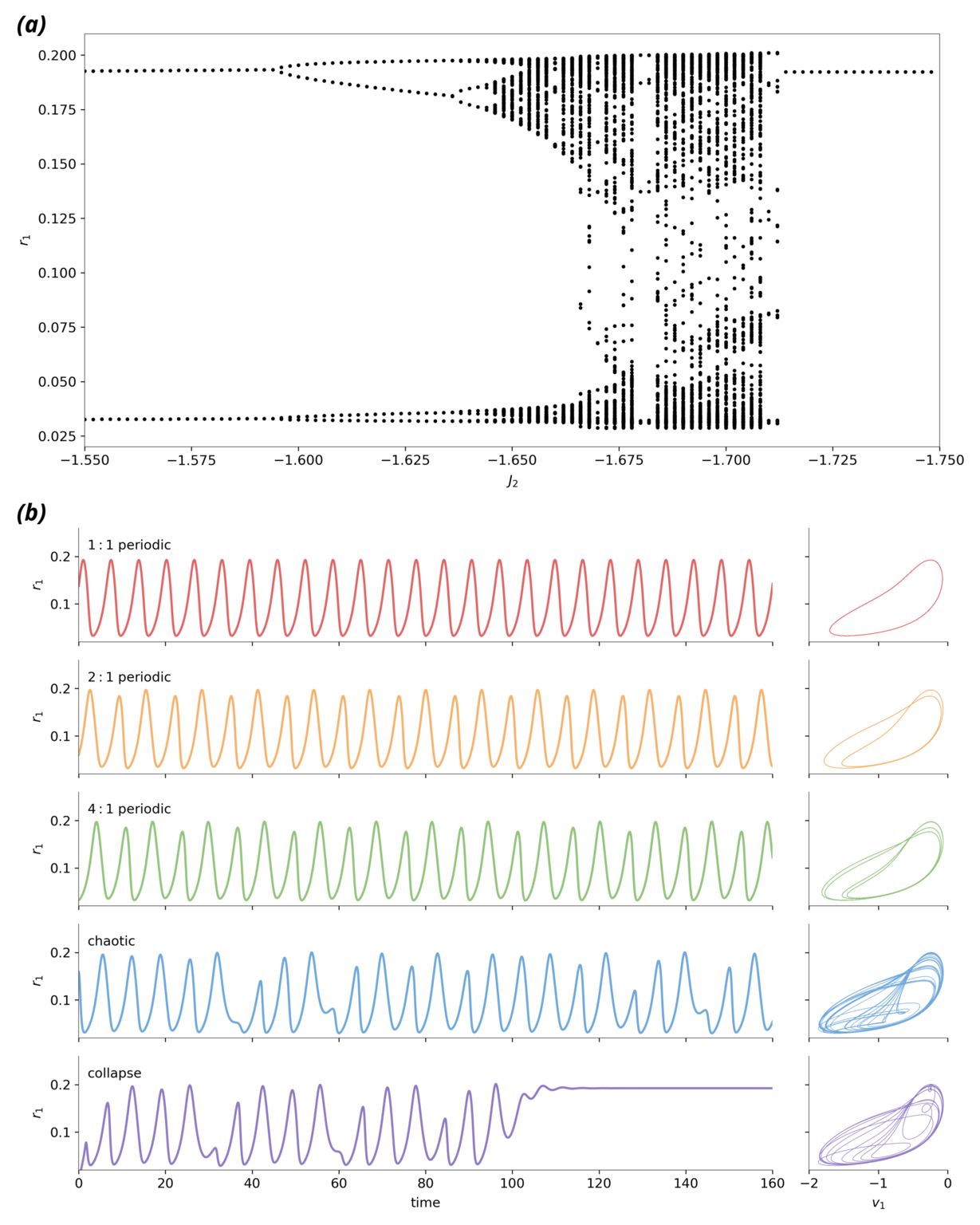}
	\captionsetup{width=.6\linewidth}
	\caption{Period-doubling cascades leading to chaos for the LC on the line $g_1=0.$ (a) Peak values of $r_1(t)$ versus $J_2$ show that as $J_2$ decreases, the LC undergoes period-doubling cascades, eventually leading to a chaotic orbit that collapses onto the other stable attractor. (b) Plots of $r_1$ versus time (left) and $r_1$ versus $v_1$ (right) illustrating the period-doubling cascades and route to chaos. From top to bottom: $J_2=-1.56, J_2=-1.62, J_2=-1.64, J_2=-1.7, J_2=-1.72.$}
	\label{fig:suppfig7}
\end{figure}

\newpage
\begin{figure}[H]
	\subfloat[SF-LC regime. Parameters: $g_1 = 0.5, J_2 = -2.5.$ ]{%
		\includegraphics[width=0.49\linewidth]{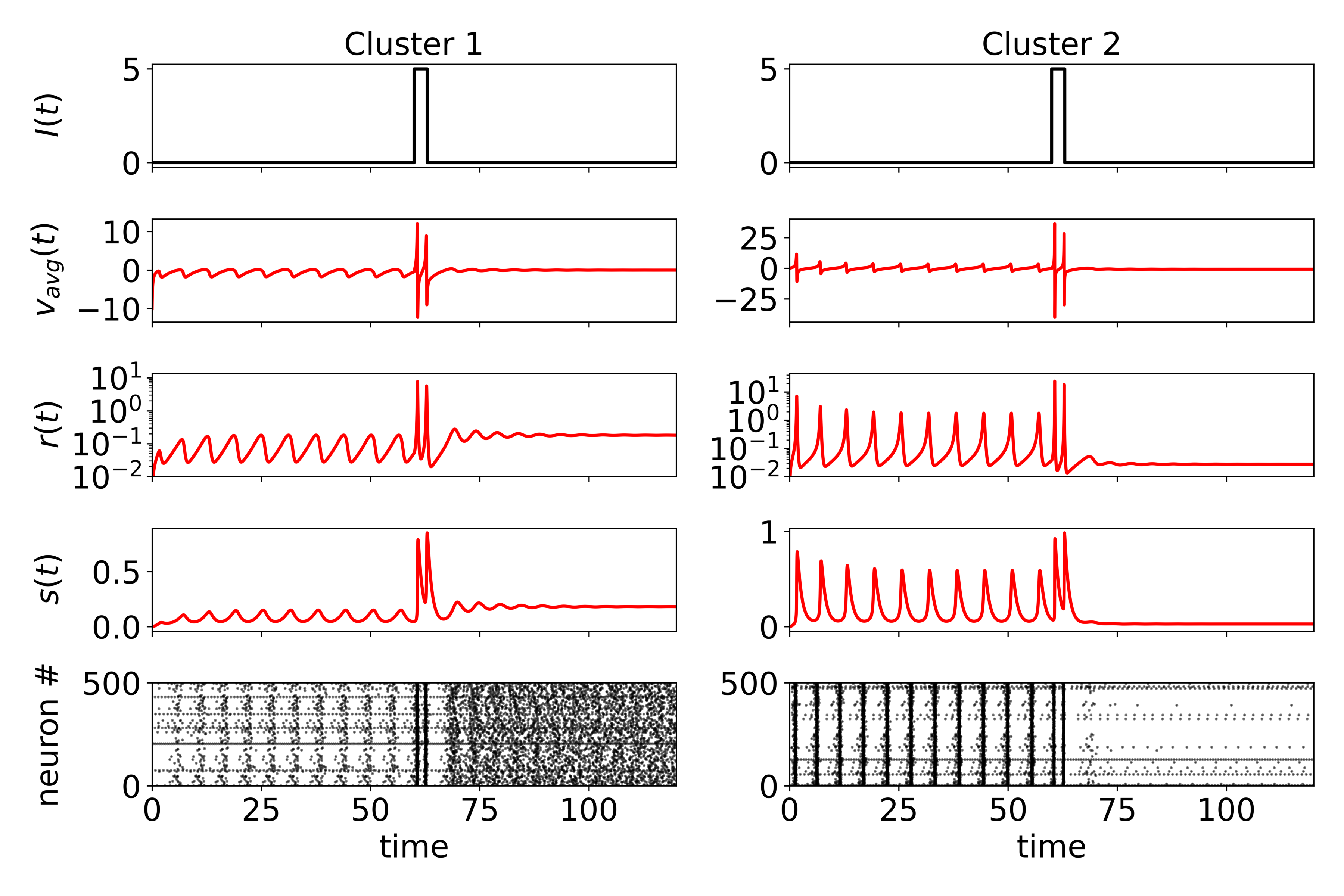}%
	}\hfill
	\subfloat[LC-LC regime. Parameters: $g_1 = 1, J_2 = -1.$ ]{%
		\includegraphics[width=0.49\linewidth]{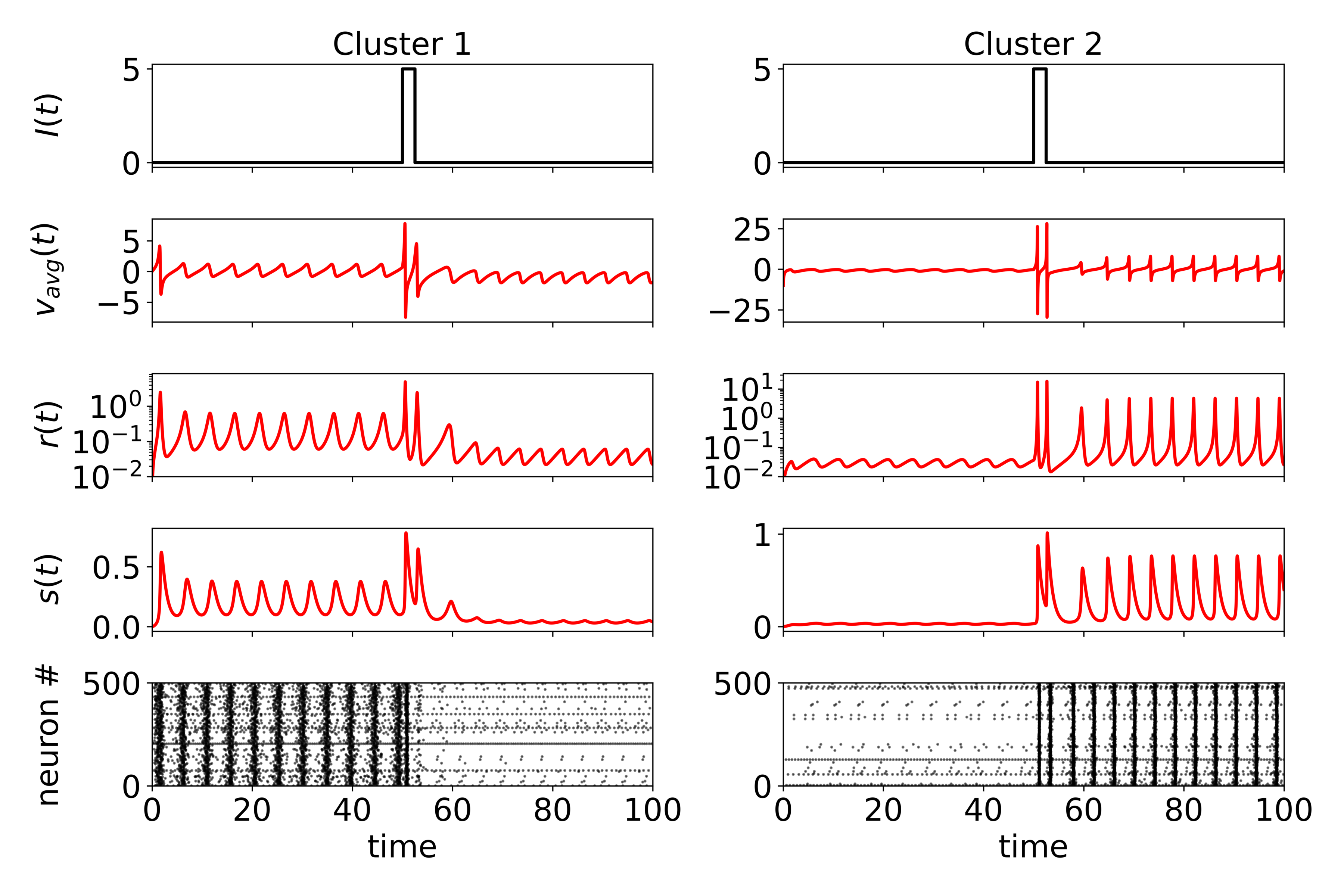}%
	}
	
	\subfloat[SF regime. Parameters: $g_1 = 0.4, J_2 = -6.$ ]{%
		\includegraphics[width=0.49\linewidth]{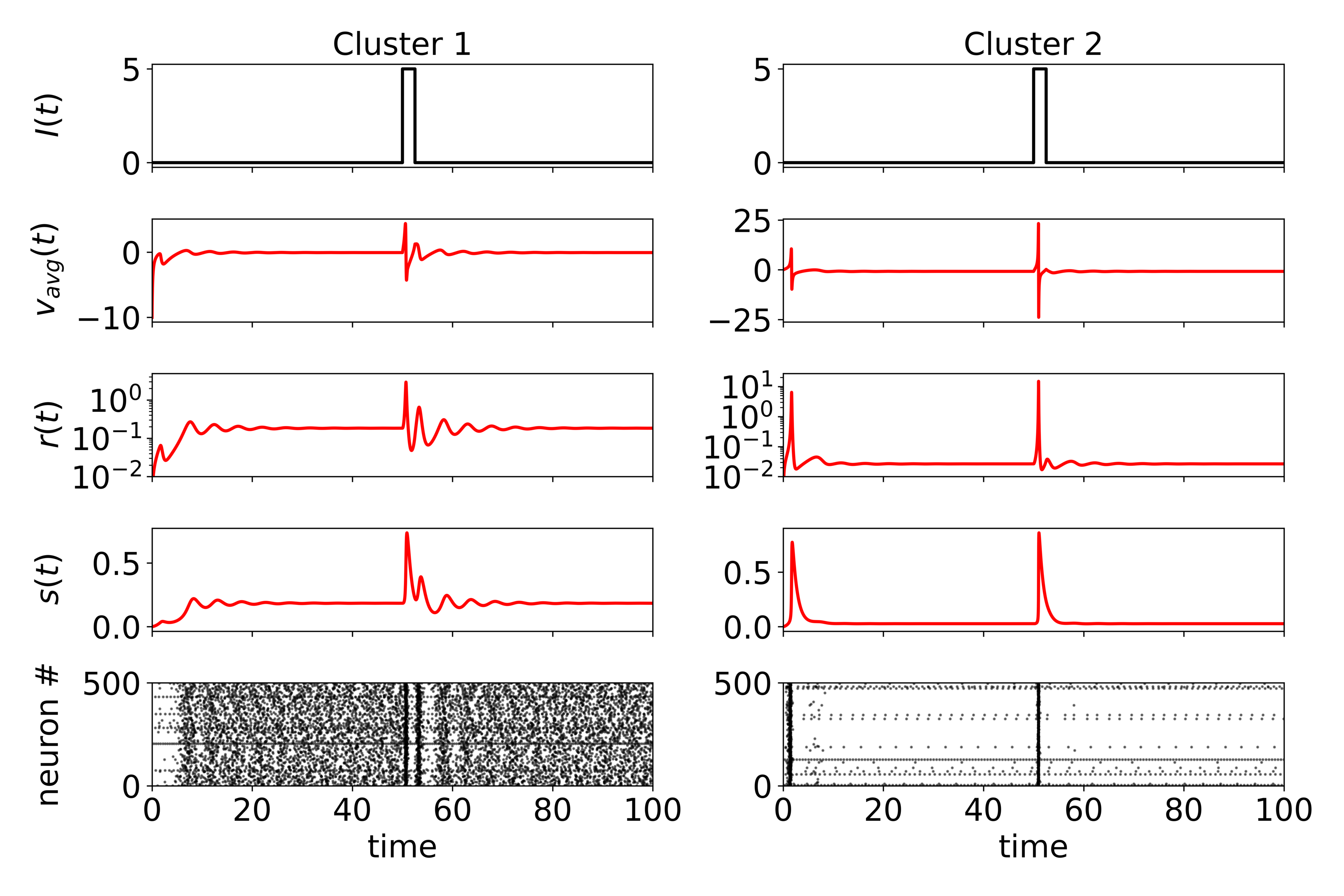}%
	}\hfill
	\subfloat[LC regime. Parameters: $g_1 = 1, J_2 = -8.$ ]{%
		\includegraphics[width=0.49\linewidth]{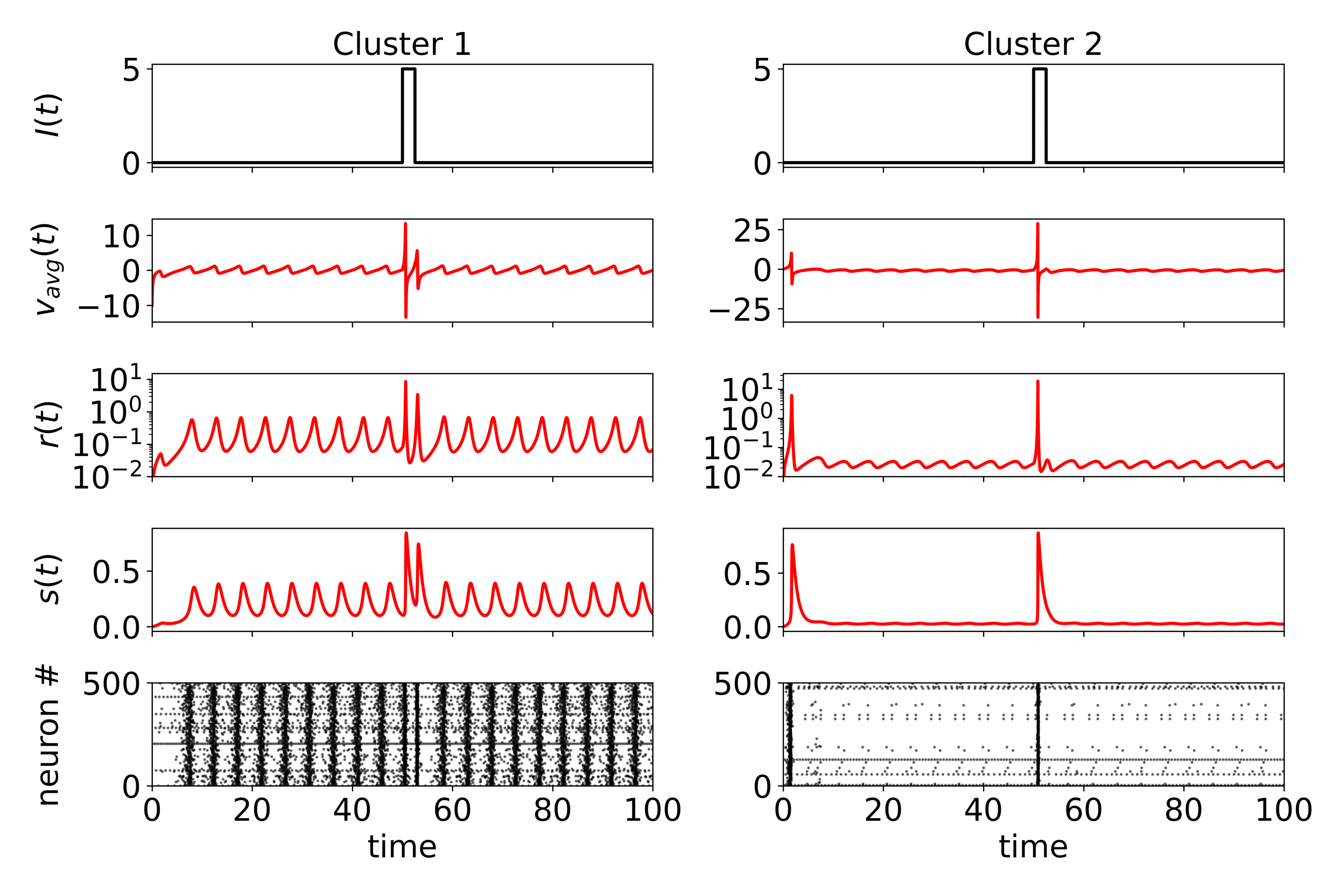}%
	}
	\captionsetup{width=\linewidth}
	\caption{Numerical simulations (black) and analytical results (red) illustrating the dynamics within each regime in Fig. 9. Parameters: $N = 10^4, g_2 = 2, J_1 = -2.5, J_c = -8, \bar{\eta}=1, \Delta=0.3$.}
	\label{fig:suppfig4} 
\end{figure}

\newpage
\begin{figure}[H]
	\subfloat[Positive pulse current injected in cluster 1. ]{%
		\includegraphics[width=0.49\linewidth]{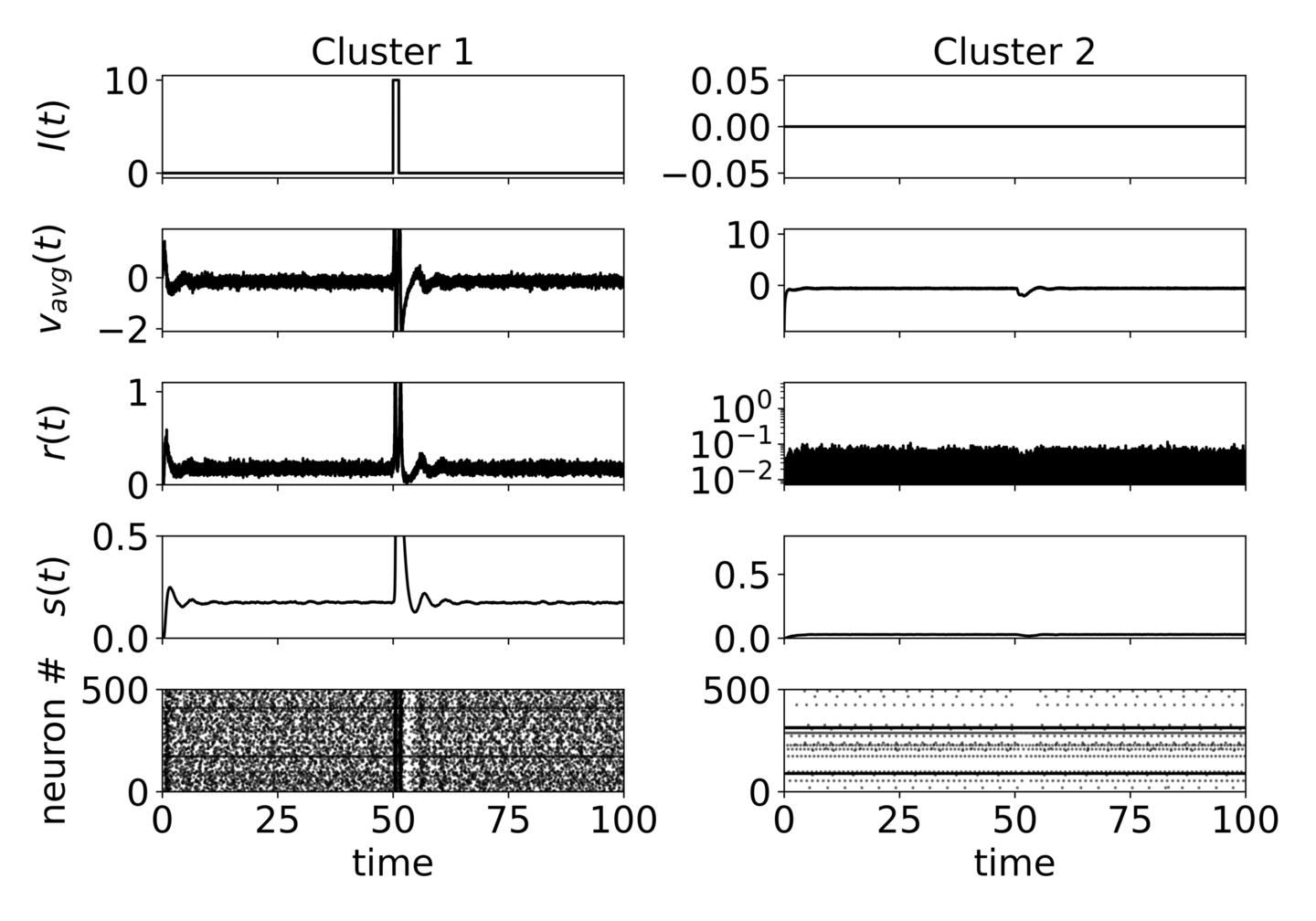}%
	}\hfill
	\subfloat[Positive pulse current injected in cluster 2. ]{%
		\includegraphics[width=0.49\linewidth]{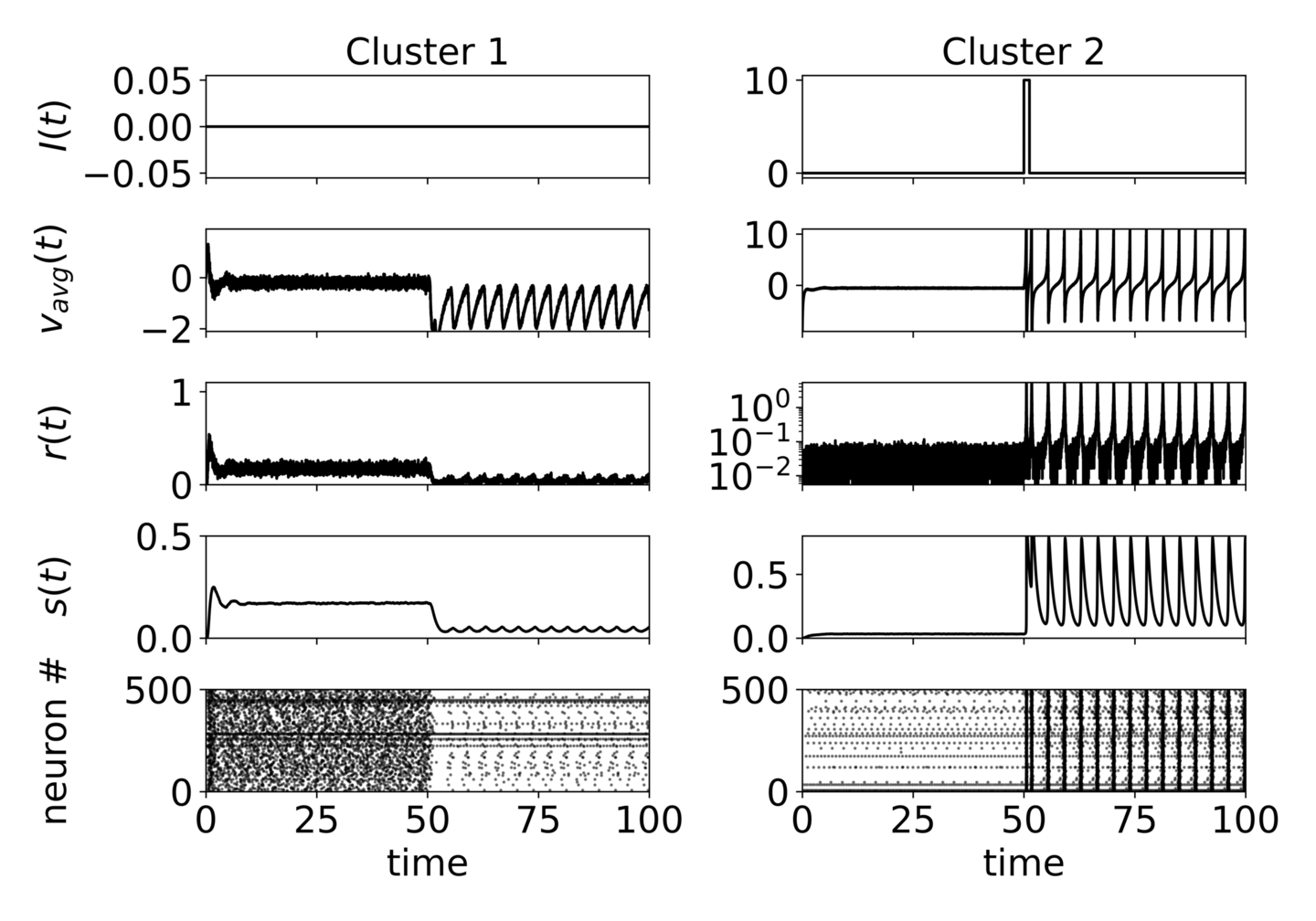}%
	}
	
	\subfloat[Negative pulse current injected in cluster 1. ]{%
		\includegraphics[width=0.49\linewidth]{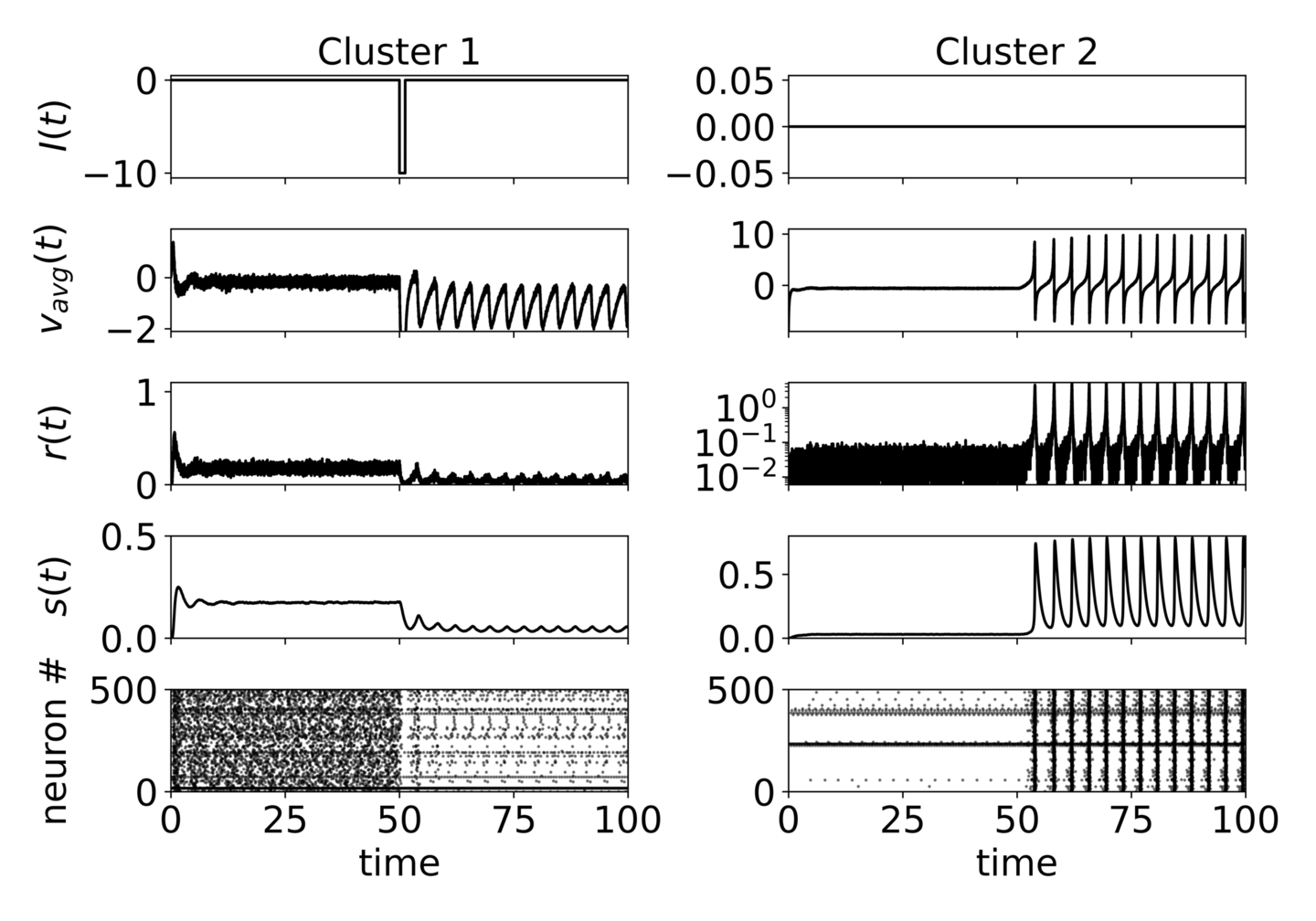}%
	}\hfill
	\subfloat[Negative pulse current injected in cluster 2. ]{%
		\includegraphics[width=0.49\linewidth]{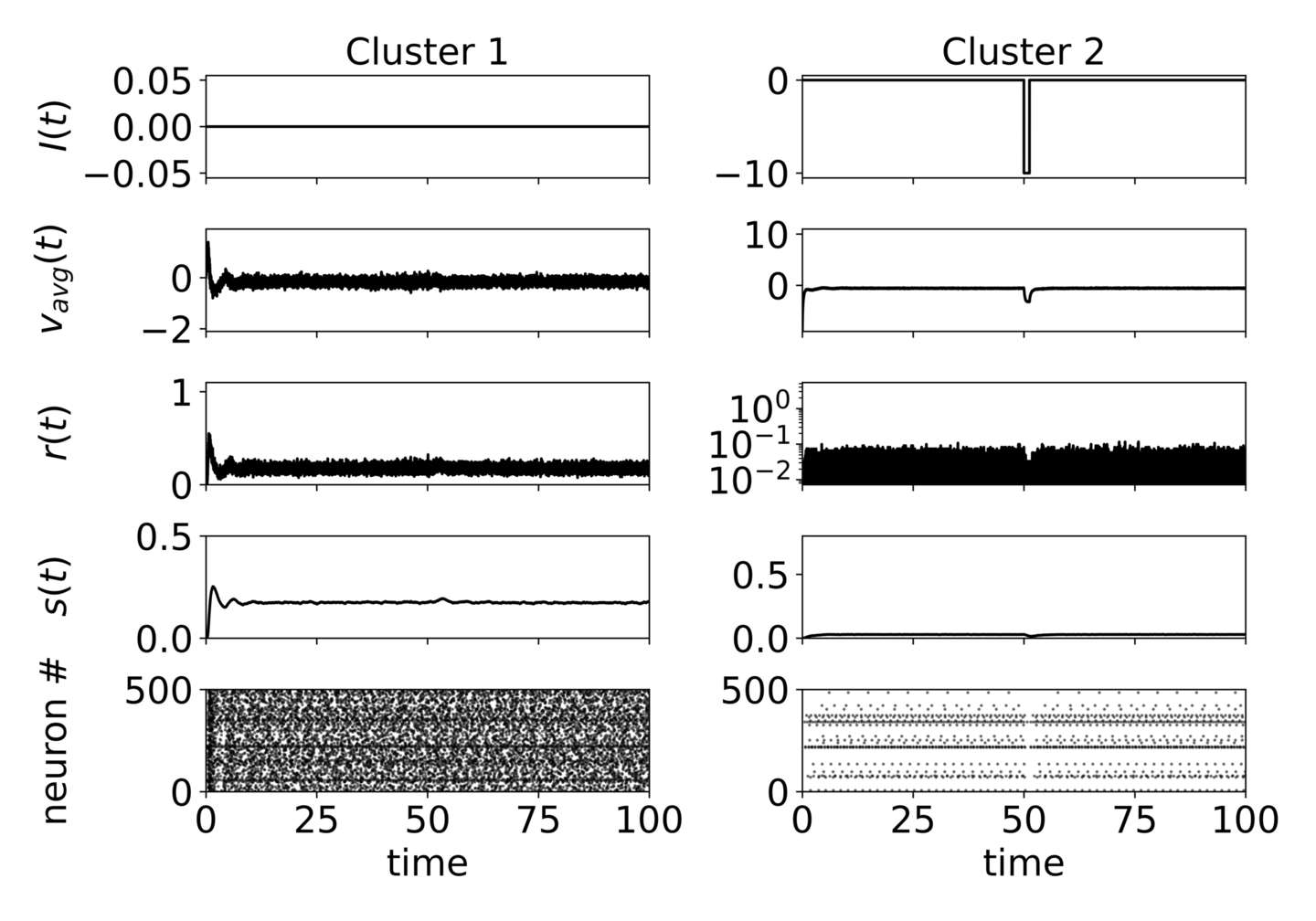}%
	}
	\captionsetup{width=\linewidth}
	\caption{Change of stability from asynchronous to synchronous regime under brief pulse current input. Neurons in cluster 1 are connected via chemical synapses and neurons in cluster 2 are connected via gap junctions. Parameters: $N = 10^4, g_1 = 0, g_2 = 2, J_1 = -2.5, J_2 = 0, J_c = -8, \bar{\eta}=1, \Delta=0.3$.}\label{fig:suppfig5}
\end{figure}

\newpage
\begin{figure}[H]
	\subfloat[Positive pulse current injected in cluster 1.]{%
		\includegraphics[width=0.49\linewidth]{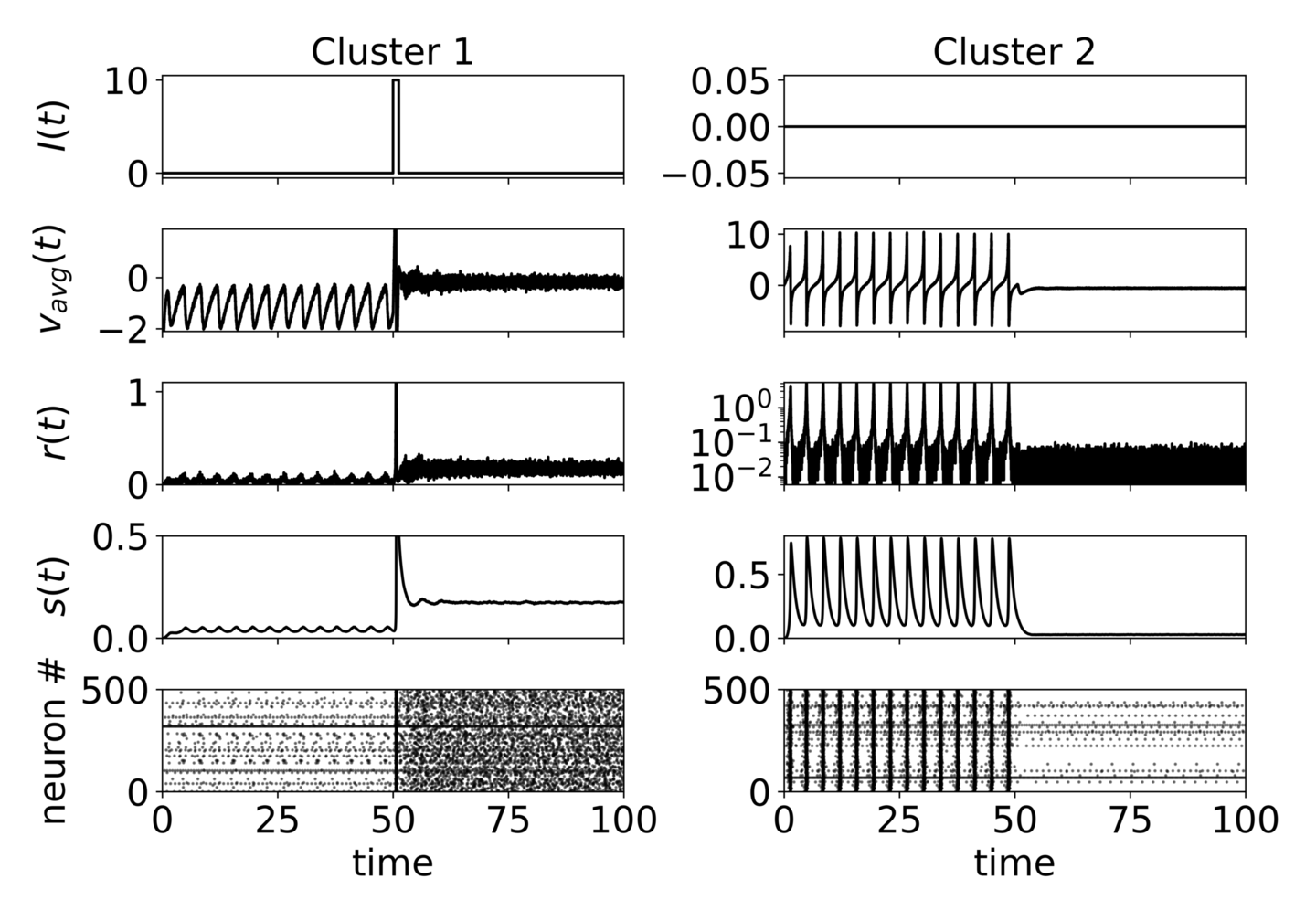}%
	}\hfill
	\subfloat[Positive pulse current injected in cluster 2.]{%
		\includegraphics[width=0.49\linewidth]{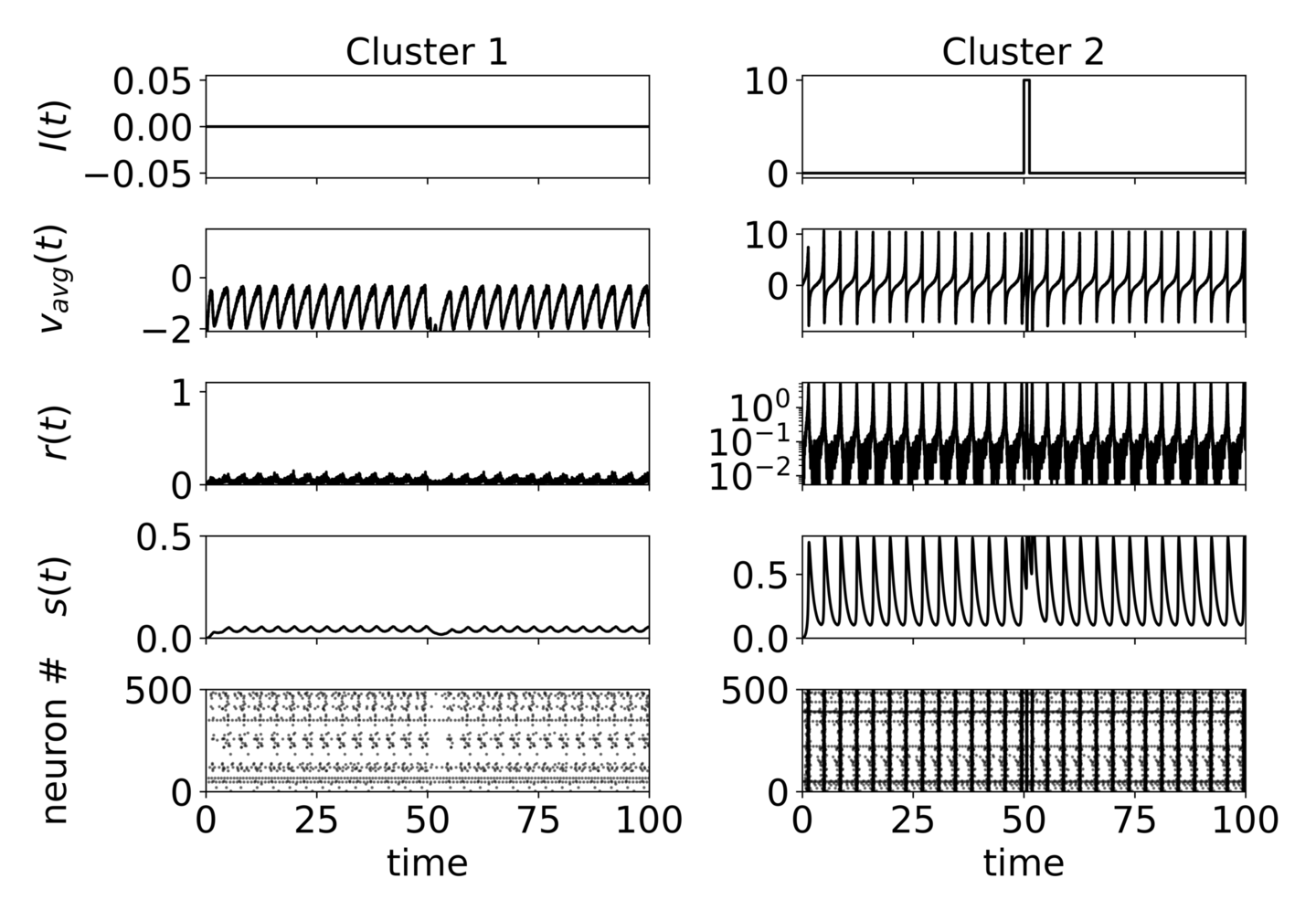}%
	}
	
	\subfloat[Negative pulse current injected in cluster 1.]{%
		\includegraphics[width=0.49\linewidth]{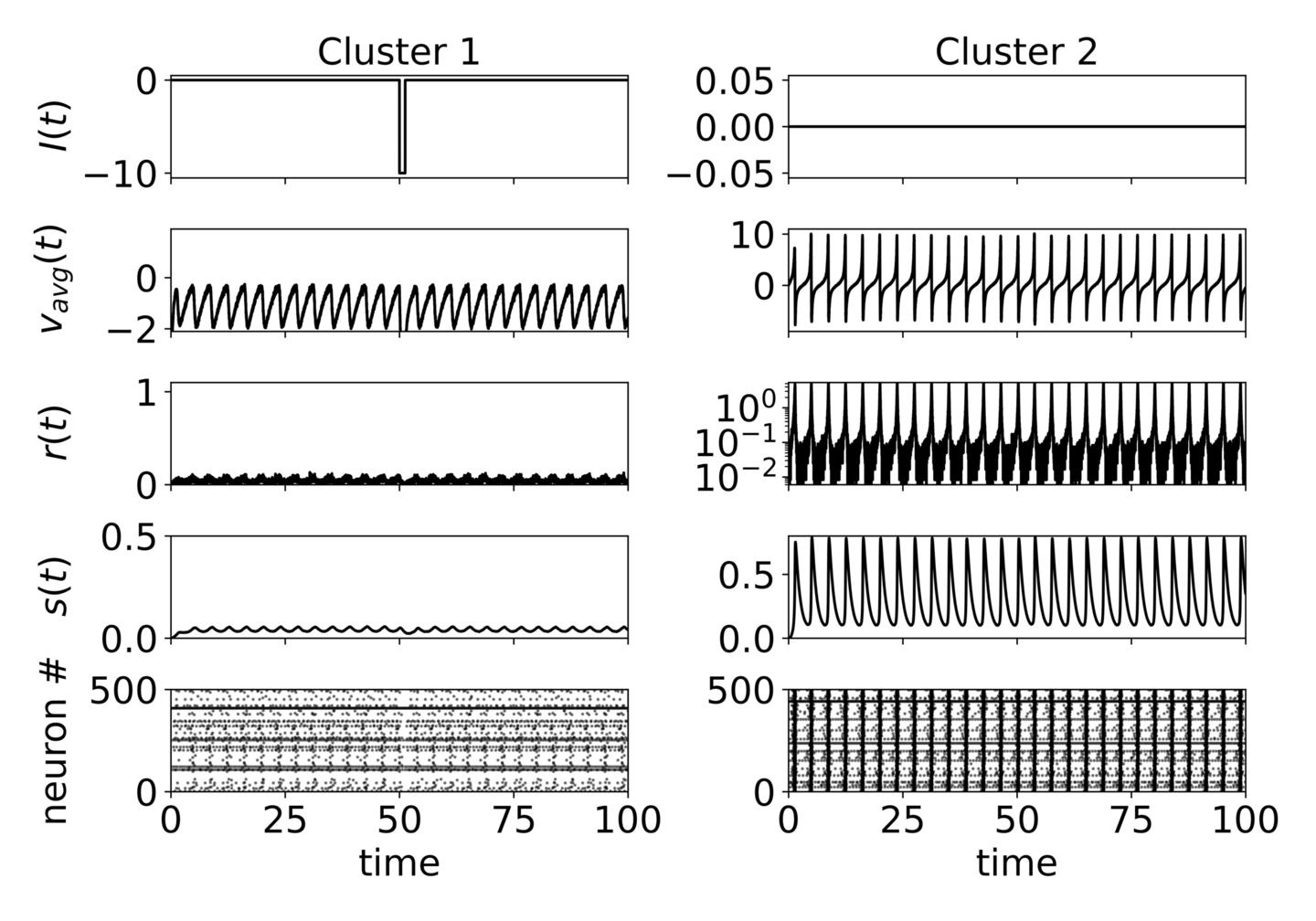}%
	}\hfill
	\subfloat[Negative pulse current injected in cluster 2.]{%
		\includegraphics[width=0.49\linewidth]{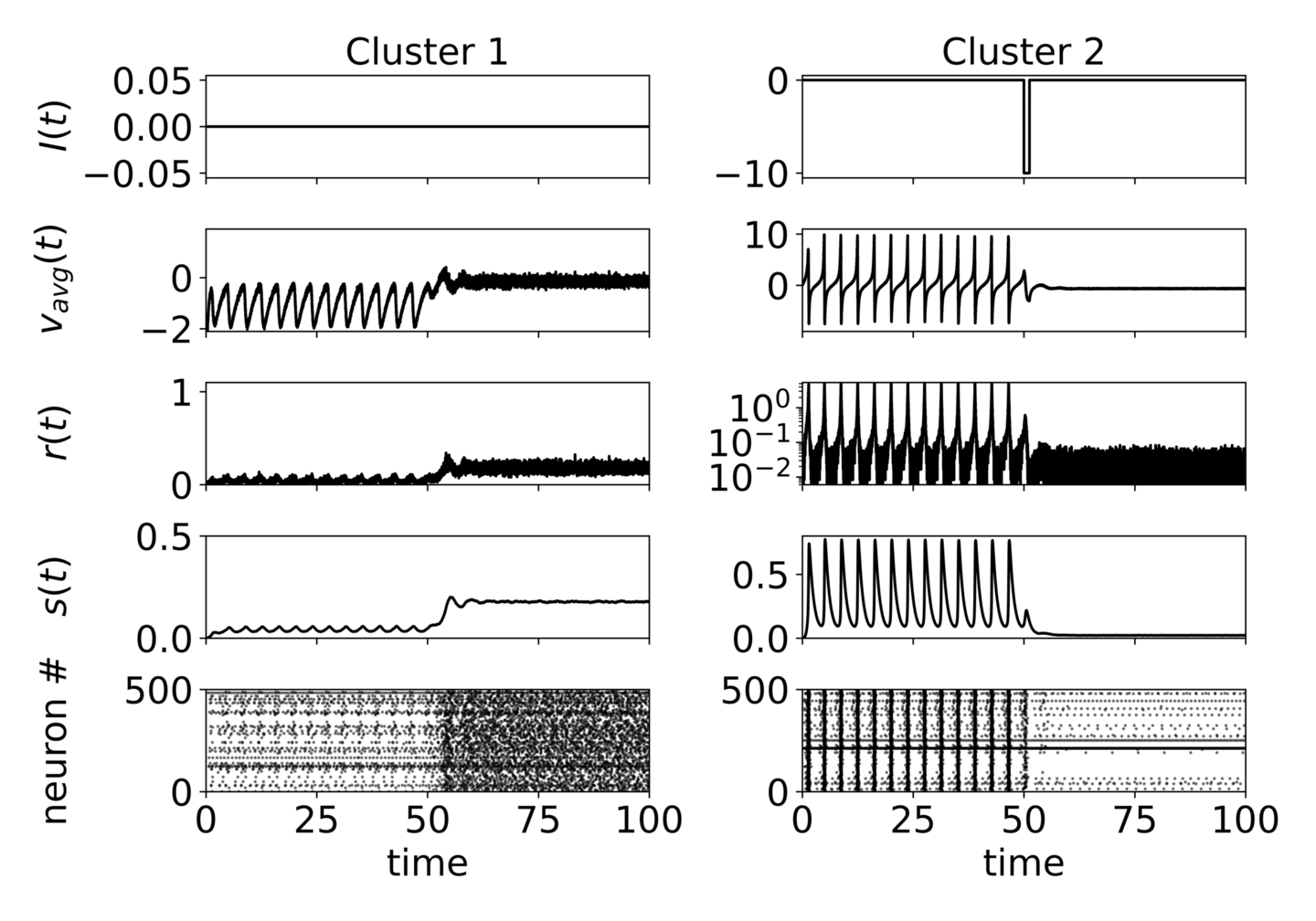}%
	}
	\captionsetup{width=\linewidth}
	\caption{Change of stability from synchronous to asynchronous regime under brief pulse current input. Neurons in cluster 1 are connected via chemical synapses and neurons in cluster 2 are connected via gap junctions. Parameters: $N = 10^4, g_1 = 0, g_2 = 2, J_1 = -2.5, J_2 = 0, J_c = -8, \bar{\eta}=1, \Delta=0.3$.} \label{fig:suppfig6}
\end{figure}

\end{document}